\let\includefigures=\iftrue
\input harvmac

% Figure definitions

\input epsf

\newcount\figno
\figno=0
\def\fig#1#2#3{
\par\begingroup\parindent=0pt\leftskip=1cm\rightskip=1cm\parindent=0pt
\baselineskip=11pt
\global\advance\figno by 1
\midinsert
\epsfxsize=#3
\centerline{\epsfbox{#2}}
\vskip 12pt
{\bf Figure \the\figno:} #1\par
\endinsert\endgroup\par
}
\def\figlabel#1{\xdef#1{\the\figno}}

% Other definitions

\noblackbox
\def\IZ{\relax\ifmmode\mathchoice
{\hbox{\cmss Z\kern-.4em Z}}{\hbox{\cmss Z\kern-.4em Z}}
{\lower.9pt\hbox{\cmsss Z\kern-.4em Z}} {\lower1.2pt\hbox{\cmsss
Z\kern-.4em Z}}\else{\cmss Z\kern-.4em Z}\fi}

\font\cmss=cmss10 \font\cmsss=cmss10 at 7pt
\def\IR{\relax{\rm I\kern-.18em R}}
\def\IN{\relax{\rm I\kern-.18em N}}

\def\frac#1#2{{#1 \over #2}}

\def\th{{\tilde h}}

%%%%%%%%%%%%%%%%%%%%%%%%%%%%%%%%%%%%%%%%%%%%%%%%%%%%%%%%%%%%%%%%%

%\draftmode
\def\journal#1&#2(#3){\unskip, \sl #1\ \bf #2 \rm(19#3) }
\def\andjournal#1&#2(#3){\sl #1~\bf #2 \rm (19#3) }

\def\d{\partial}

%
%%%%%%%%%%%%%%%%%%%%%%%%%%%%%%%%%%%%
%

%
\catcode`\@=11
\def\slash#1{\mathord{\mathpalette\c@ncel{#1}}}
\overfullrule=0pt

\def\HH{{\cal H}}

\def\TT{{\cal T}}

\def\underrel#1\over#2{\mathrel{\mathop{\kern\z@#1}\limits_{#2}}}

\catcode`\@=12

\def\({\left(}
\def\){\right)}
\def\[{\left[}
\def\]{\right]}

\def\d{\partial}

\def \sinh{{\rm sinh}}
\def \cosh{{\rm cosh}}
\def \tanh{{\rm tanh}}
\def \coth{{\rm coth}}

\def\exp{{\rm exp}}
\def\sh{{\rm sinh}}
\def\ch{{\rm cosh}}
\def \th{{\rm tanh}}

%%%%%%%%%%%%%%%%%%%%%%%%%%%%%%%%%%%%%%%%%%%%%%%%%%%%%%%%%%%%%%

%

%%%%%%%%%%%%%%%%%%%%%%%%%%%%%%%%%%%%%%%%%%%%%%%%%%%%%%%%%%%%%%
% new defs:

%\def\myTitle#1#2{\nopagenumbers\abstractfont\hsize=\hstitle\rightline{#1}%
%\vskip 0.5in\centerline{\titlefont #2}\abstractfont\vskip .5in\pageno=0}
%
%\myTitle{\vbox{\baselineskip12pt\hbox{} \hbox{RI-04-04}}}

\rightline{RI-04-04}
\Title{
\rightline{hep-th/0406131}}
{\vbox{\centerline{On Thermodynamical Properties of Some}
        \centerline{Coset CFT Backgrounds}
    \medskip    }}
\centerline{\it Amit~Giveon, Anatoly~Konechny, Eliezer~Rabinovici and
Amit~Sever\foot{E-mails : {\tt giveon@phys.huji.ac.il,
tolya@phys.huji.ac.il, eliezer@vms.huji.ac.il,
asever@phys.huji.ac.il}} }
\medskip
\centerline{Racah Institute of Physics, The Hebrew University,
Jerusalem 91904, Israel}

\bigskip
\bigskip

\noindent
We investigate the thermodynamical features of two Lorentzian
signature backgrounds that arise in string theory as exact CFTs
and possess more than two disconnected asymptotic regions: the 2-d
charged black hole and the Nappi-Witten cosmological model. We
find multiple smooth disconnected  Euclidean versions of the
charged black hole background. They are characterized by different
temperatures and electro-chemical potentials. We show that there
is no straightforward analog of the Hartle-Hawking state that
would express these thermodynamical features. We also obtain
multiple Euclidean versions of the Nappi-Witten cosmological
model and study their singularity structure. It suggests to
associate a non-isotropic temperature with this background.

\vfill

\Date{}

\lref\BBBC{
S.~Elitzur, A.~Giveon, D.~Kutasov and E.~Rabinovici,
``From big bang to big crunch and beyond,''
JHEP {\bf 0206}, 017 (2002)
[arXiv:hep-th/0204189].
%%CITATION = HEP-TH 0204189;%%
}

\lref\removing{
S.~Elitzur, A.~Giveon and E.~Rabinovici,
``Removing singularities,''
JHEP {\bf 0301}, 017 (2003)
[arXiv:hep-th/0212242].
%%CITATION = HEP-TH 0212242;%%
}

\lref\ChargedBh{
A.~Giveon, E.~Rabinovici and A.~Sever,
``Beyond the singularity of the 2-D charged black hole,''
JHEP {\bf 0307}, 055 (2003)
[arXiv:hep-th/0305140].
%%CITATION = HEP-TH 0305140;%%
}

\lref\Israel{W.~Israel, {\it Thermo-field dynamics of black holes
}, Phys. Lett. {\bf 57A} 107 (1976).}

\lref\Waldbook{R.~M.~Wald, {\it Quantum field theory in curved
spacetime and black hole thermodynamics}, Univ. of Chicago press,
1994.}

\lref\NW{C.~R.~Nappi and E.~Witten, {\it A closed, expanding
universe in string theory}, Phys.Lett. {\bf B293} (1992) 309-314;
hep-th/9206078.}

\lref\McGNY{M.~D.~McGuigan, C.~R.~Nappi and S.~A.~Yost, {\it
Charged black holes in two-dimensional string theory}, Nucl. Phys.
{\bf B375}, 421 (1992); hep-th/9111038.}

\lref\ILS{N.~Ishibashi, M.~Li and A.~R.~Steif, {\it
Two-dimensional charged black holes in string theory}, Phys. Rev.
Lett. {\bf 67}, 3336 (1991).}

\lref\GP{G.~W.~Gibbons and M.~J.~Perry, {\it The physics of 2-d
stringy spacetimes}, Int. J. Mod. Phys. {\bf D1} (1992) 335-354;
hep-th/9204090.}

\lref\NP{C.~R.~Nappi and A.~Pasquinucci, {\it Thermodynamics of
two-dimensional black holes}, Mod. Phys. Lett. {\bf A7} (1992)
3337-3346; gr-qc/9208002.}

\lref\BP{B.~Pioline and M.~Berkooz, {\it Strings in an electric
field, and the Miln Universe}, hep-th/0307280.}

\lref\Vil{N.~J.~Vilenkin, {\it Special functions and the theory of
group representations}, AMS, 1968.}

\lref\Maldacena{J.~Maldacena, {\it  Eternal black holes in AdS},
JHEP {\bf 0304} (2003) 021; hep-th/0106112.}

\lref\gkos{A.~Giveon, B.~Kol, A.~Ori and A.~Sever, ``On the
resolution of the time-like singularities in Reissner-Nordstroem
and negative-mass Schwarzschild,'' arXiv:hep-th/0401209.}

\lref\dvv{ R.~Dijkgraaf, H.~Verlinde and E.~Verlinde, ``String
propagation in a black hole geometry,'' Nucl.\ Phys.\ B {\bf 371},
269 (1992).}

\lref\witten{ E.~Witten, ``On string theory and black holes,''
Phys.\ Rev.\ D {\bf 44}, 314 (1991).}

\lref\DIsrael{D.~Israel, C.~Kounnas and M.~P.~Petropoulos,
``Superstrings on NS5 backgrounds, deformed AdS(3) and
holography,'' JHEP {\bf 0310} (2003) 028 [arXiv:hep-th/0306053].}

\lref\Wald{ A.~Ishibashi and R.~M.~Wald, ``Dynamics in
non-globally hyperbolic static spacetimes. III: anti-de Sitter
spacetime,'' arXiv:hep-th/0402184. A.~Ishibashi and R.~M.~Wald,
``Dynamics in non-globally-hyperbolic static spacetimes. II:
General  analysis of prescriptions for dynamics,'' Class.\ Quant.\
Grav.\  {\bf 20}, 3815 (2003) [arXiv:gr-qc/0305012]. R.~M.~Wald,
``Dynamics In Nonglobally Hyperbolic, Static Space-Times,'' J.\
Math.\ Phys.\  {\bf 21}, 2802 (1980).}

\lref\Elitzur{ S.~Elitzur, A.~Forge and E.~Rabinovici, ``Some
Global Aspects Of String Compactifications,'' Nucl.\ Phys.\ B {\bf
359}, 581 (1991). G.~Mandal, A.~M.~Sengupta and S.~R.~Wadia,
``Classical solutions of two-dimensional string theory,'' Mod.\
Phys.\ Lett.\ A {\bf 6}, 1685 (1991).}

\lref\Unruh{ W.~G.~Unruh, ``Notes On Black Hole Evaporation,''
Phys.\ Rev.\ D {\bf 14}, 870 (1976).}

\lref\giv{ A.~Giveon, ``Target space duality and stringy black
holes,'' Mod.\ Phys.\ Lett.\ A {\bf 6}, 2843 (1991).}

\lref\GPR{A.~Giveon, M.~Porrati and E.~Rabinovici, ``Target space
duality in string theory,'' Phys.\ Rept.\  {\bf 244}, 77 (1994)
[arXiv:hep-th/9401139].}

\lref\brs{ K.~Bardakci, E.~Rabinovici and B.~Saering, ``String
Models With $C<1$ Components,'' Nucl.\ Phys.\ B {\bf 299}, 151
(1988); D.~Altschuler, K.~Bardakci and E.~Rabinovici, ``A
Construction Of The $C<1$ Modular Invariant Partition Functions,''
Commun.\ Math.\ Phys.\  {\bf 118}, 241 (1988); W.~Nahm, ``Gauging
Symmetries Of Two-Dimensional Conformally Invariant Models,''
UCD-88-02; K.~Gawedzki and A.~Kupiainen, ``Coset Construction From
Functional Integrals,'' Nucl.\ Phys.\ B {\bf 320}, 625 (1989);
D.~Karabali, Q.~H.~Park, H.~J.~Schnitzer and Z.~Yang, ``A Gko
Construction Based On A Path Integral Formulation Of Gauged
Wess-Zumino-Witten Actions,'' Phys.\ Lett.\ B {\bf 216}, 307
(1989).}

\lref\Gibbons{ G.~W.~Gibbons and S.~W.~Hawking, ``Action Integrals
And Partition Functions In Quantum Gravity,'' Phys.\ Rev.\ D {\bf
15}, 2752 (1977).}

\lref\KayWald{ B.~S.~Kay and R.~M.~Wald, ``Theorems On The
Uniqueness And Thermal Properties Of Stationary, Nonsingular,
Quasifree States On Space-Times With A Bifurcate Killing
Horizon,'' Phys.\ Rept.\  {\bf 207}, 49 (1991); B.~S.~Kay,
``Sufficient Conditions For Quasifree States And An Improved
Uniqueness Theorem For Quantum Fields On Space-Times With
Horizons,'' J.\ Math.\ Phys.\  {\bf 34}, 4519 (1993). }

\lref\BBW{ H.~W.~Braden, J.~D.~Brown, B.~F.~Whiting and
J.~W.~.~York, ``Charged Black Hole In A Grand Canonical
Ensemble,'' Phys.\ Rev.\ D {\bf 42}, 3376 (1990).}

\lref\Penrose{R. Penrose, in Battelle Rencontres, 1967 lectures in
mathematics and physics , edited by C. M. DeWitt and J. A. Wheeler
(Benjamin, New York, 1968), P. 222 . E.~Poisson and W.~Israel,
``Internal Structure Of Black Holes,'' Phys.\ Rev.\ D {\bf 41},
1796 (1990) and references therein.}

\lref\Ori{A. Ori, ``Inner structure of a charged black hole: An
exact mass-inflation solution,'' Phys. Rev. Lett. 67 , 789 (1991).
L.~M.~Burko, ``Structure of the black hole's Cauchy horizon
singularity,'' Phys.\ Rev.\ Lett.\  {\bf 79}, 4958 (1997)
[arXiv:gr-qc/9710112]. This is also the situation in the more
realistic, spinning black-hole case: see e.g. A. Ori, ``Structure
of the singularity inside a realistic rotating black hole,'' Phys.
Rev. Lett. 68 , 2117 (1992); A.~Ori, ``Evolution Of Linear
Gravitational And Electromagnetic Perturbations Inside A Kerr
Black Hole,'' Phys.\ Rev.\ D {\bf 61}, 024001 (2000).}

\lref\VW{ C.~Vaz and L.~Witten, ``The quantum states and the
statistical entropy of the charged black  hole,'' Phys.\ Rev.\ D
{\bf 63}, 024008 (2001) [arXiv:gr-qc/0006039].}

\lref\HawHll{S.~Hawking and G.~Ellis, "The large scale structure
of space-time," Cambridge University Press,
1973.}

\lref\Shenker{ P.~Kraus, H.~Ooguri and S.~Shenker, ``Inside the
horizon with AdS/CFT,'' Phys.\ Rev.\ D {\bf 67}, 124022 (2003)
[arXiv:hep-th/0212277]. L.~Fidkowski, V.~Hubeny, M.~Kleban and
S.~Shenker, ``The black hole singularity in AdS/CFT,'' JHEP {\bf
0402}, 014 (2004) [arXiv:hep-th/0306170].}

\lref\Seiberg{ H.~Liu, G.~Moore and N.~Seiberg, ``Strings in a
time-dependent orbifold,'' JHEP {\bf 0206}, 045 (2002)
[arXiv:hep-th/0204168]. H.~Liu, G.~Moore and N.~Seiberg,
%``Strings in time-dependent orbifolds,''
JHEP {\bf 0210}, 031 (2002) [arXiv:hep-th/0206182].}

\lref\Hiscock{ W.~A.~Hiscock, ``Stress Energy Tensor For A
Two-Dimensional Evaporating Black Hole,'' Phys.\ Rev.\ D {\bf 16},
2673 (1977).}

\lref\Horowitz{ G.~T.~Horowitz and J.~Polchinski, ``Instability of
spacelike and null orbifold singularities,'' Phys.\ Rev.\ D {\bf
66}, 103512 (2002) [arXiv:hep-th/0206228].}

\lref\proc{ A.~Giveon, E.~Rabinovici and A.~Sever, ``Strings in
singular time-dependent backgrounds,'' Fortsch.\ Phys.\  {\bf 51},
805 (2003) [arXiv:hep-th/0305137]; E.~Rabinovici, ``Aspects of
string motion in time-dependent backgrounds,'' Annales Henri
Poincare {\bf 4}, S159 (2003). }

\lref\Lee{ H.~W.~Lee, Y.~S.~Myung and J.~Y.~Kim, ``Blushift of
tachyon in the charged 2D black hole,'' Phys.\ Rev.\ D {\bf 52},
5806 (1995) [arXiv:hep-th/9510122].}

\lref\Hull{ C.~M.~Hull, ``Timelike T-duality, de Sitter space,
large N gauge theories and  topological field theory,'' JHEP {\bf
9807}, 021 (1998) [arXiv:hep-th/9806146].}

\newsec{Introduction} General Relativity (GR) as a low energy
theory of gravity suffers from several problems. Perhaps some of
these problems can be resolved in a more complete theory of
gravity. String theory is an attempt to construct such a theory.
At low energies string theory should however
reduce in many ways back to GR and thus may well face itself those
problem plaguing low energy gravity. It seems that if string
theory does play a role in fixing problems in gravity it will do
it by adding a ``twist'' to GR -- a new point of view  that could have
been in principle envisaged already in a GR framework.

The addition of massless particles motivated by string theory does
help in resolving some of the central problems in GR, such as those related
to the presence of singularities. Another manner in which
string theory can influence GR is by enlarging the portion of 
space-time arena which one would need to probe. For example, the
Lorentzian uncharged black-hole
string background contains two disconnected static regions,
which we shall somewhat loosely call boundaries.
%~\foot{For 
%a black hole in Anti de-Sitter space-time they are actually boundaries.}.
There is also a region between the horizons and the singularities.
A charged black hole (CBH) in GR has an inner horizon, an outer
horizon and a time-like singularity \HawHll. A CBH obtained from
string theory has in addition asymptotic regions that extend
beyond its time-like singularities \ChargedBh. A similar
enlargement occurs in the case of exact string backgrounds that
describe a closed cosmology. There one is led in string theory  to
a picture in which the closed cosmology is embedded within static
non-compact regions called ``whiskers'' \BBBC. 

What is common to
all these Lorentzian examples is that they have two or more
asymptotic regions \refs{\BBBC,\ChargedBh,\removing}. In string theory it
is natural to study these extended space-time regions. It turns
out that one can calculate the scattering matrix for waves to
pass from the whiskers through the big bang/big crunch space-like 
singularities \refs{\BBBC,\removing}
as well as beyond the time-like CBH singularities
\refs{\ChargedBh,\gkos}. 
All these calculations do not take into
account the possible back reaction which in other cases have
shown to be fatal \refs{\Seiberg,\Horowitz} 
(for a more complete list of references see \proc). 
In fact, one usually expects that back reaction
will eventually detach the regions of space-time beyond a big
crunch singularity as well as those regions beyond a CBH's
singularity. Actually, one would expect already the regions
beyond its inner horizon to decouple \refs{\Penrose,\Lee}; see
however \Ori.

String theory could assist in evaluating the actual role of back
reaction if/once a holographic description is found. In its
absence, another way to assess the importance of back reaction
could be addressed by considering the appropriate Euclidean version of the
system \Shenker. Once the propagation on such a background is
found to be free of singularities, one may investigate the manner
in which the more elaborate Lorentzian structure is encoded in the
Euclidean one. One may also attempt to determine the extent to
which it can be used to detect effects of singularities.

In the case of an uncharged AdS black hole it has been shown that
the effect of having two Lorentzian boundaries is equivalent to
having a system with an entangled pure state \refs{\Israel,\Maldacena}.
This is encoded in the Euclidean version through the emergence of
thermodynamical properties such as the temperature of the
black hole. Such properties are not associated with a
usual vacuum state. 

In this note we wish to investigate the
thermodynamical properties of Euclidean versions of Lorentzian systems 
that have more than two asymptotic regions. It could
be that in the presence of horizons such cases can be described by
pure entangled states. It may also turn out that such systems
posses more than one Euclidean version. 

In section 2 we discuss
the Euclidean versions of two-dimensional CBHs corresponding to
exact quotient CFTs. These systems have
various boundaries separated by both an outer (event) and an inner
(Cauchy) horizon. We find them to have two Euclidean versions,
each with its own associated temperature \Hiscock, specific heat
and electro-chemical potential. 
The electro-chemical potential is related
to the value of a Wilson loop.
We further discuss a Lorentzian
derivation of the temperatures associated with the two different
Euclidean versions. 

In section 3 we investigate the
quantization of a scalar field in the extended 2-d CBH background.
We describe an attempt to find a simple generalization of the
Hartle-Hawking state for the extended CBH -- a system which a priori has at
least four asymptotic regions. We show, under certain assumptions,
that a class of such states does not exist. 

In section 4 we
discuss the Euclidean version of a cosmological example in which
there are several Lorentzian boundaries separated by big bang/big
crunch singularities and not by horizons \refs{\NW,\BBBC}. This is also an
exact coset CFT which describes a cosmology with whiskers. We
obtain its Euclidean version by analytic continuation, and discuss
its singularity structure as well as possible
non-isotropic thermodynamical properties. The same answer is
also obtained by doing a Euclidean coset.

In appendix A we extend the discussion of the
relation between the value of a
Wilson loop and the value of the chemical potential
conjugate to angular momentum in Kerr black holes.
In appendices B, C we
elaborate on the tools used in the 
discussion of quantization of a scalar field in the CBH background.

\newsec{Charged 2-d black holes}
\subsec{Lorentzian charged 2-d black hole -- a review}

A class of charged 2-d black hole backgrounds arising from coset
CFT's was recently studied in \ChargedBh. One begins with a
three-dimensional background arising by gauging \brs\ a WZW model 
$SL(2,{\IR})\times U(1)$ by a non-compact $U(1)$ subgroup. The
gauge group  action can be chosen as \ChargedBh
 \eqn\gaugecbh{
    (g, x_{L}, x_{R})\mapsto(e^{\rho\sigma_{3}/\sqrt{k}}
    ge^{\tau\sigma_{3}/\sqrt{k}},x_{L}+\rho', x_{R} + \tau').}
Here $g\in SL(2, \IR)$, $x\in U(1)$ and $k>0$ is the $SL(2, \IR)$
level. Picking a single non-anomalous $U(1)$ gauge group can then
be achieved by imposing the constraints
 \eqn\taurho{\underline{\rho} = R\underline \tau~,}
where $\underline{\tau}=(\tau,\tau')$, $\underline \rho=(\rho,
\rho')$, and $R$ is an $SO(2)$ matrix
 \eqn\rrr{
    R=\(\matrix{\cos(\psi)& \sin(\psi)\cr
    -\sin(\psi) &\cos(\psi)}\) \, .}
In addition to \taurho\ we set
 \eqn\taufix{\underline{\tau}= |\tau|(1,0)~.}
Gauging with respect to the so defined $U(1)$ defines a family of
coset CFT's $\frac{SL(2,\IR)\times U(1)}{U(1)}$ labelled by the
angular parameter $\psi$. The metric and dilaton backgrounds can
be obtained from the gauged WZW model action by fixing the gauge
and then integrating the gauge fields out in the path integral.
One can further perform a Kaluza-Klein (KK) reduction
along the $x$-direction using the facts that the $x$ direction can
be chosen relatively very small and that in the large $k$ limit,
which is of interest to us, low energy gravity is a good
approximation. Upon the KK reduction one obtains a
background describing a two-dimensional CBH. The corresponding
Penrose diagram contains six basic regions that are periodically
repeated in the maximally extended solution.

 \fig{Penrose diagram of the 2-d charged black hole.}{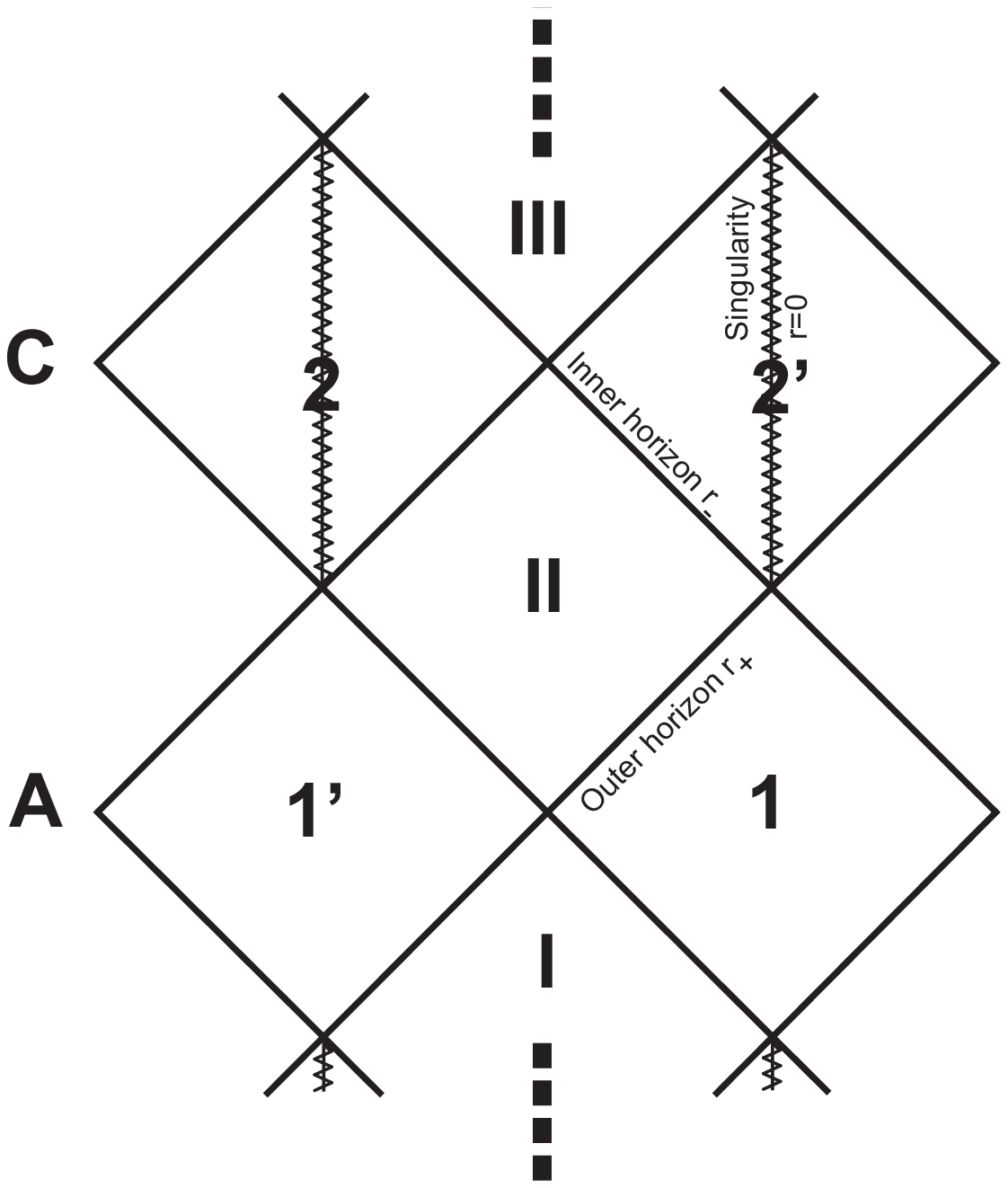}
 {6 truecm} \figlabel\cbh

We will concentrate on the two static regions that are analogs of
the regions outside the outer and and beyond the inner horizons in the
four-dimensional Reissner-Nordstrom (RN) solution. A smooth region, to
which we will refer to as a region of type A, is represented on
the Penrose diagram (Fig. 1) by the union of regions 1 and 1'
while a region containing the singularity, to be referred to as
type C region, is represented by the union of regions 2 and 2'. We
parameterize the corresponding regions in $SL(2,\IR)$ by
 \eqn\slparam{
    g_A=e^{{1\over 2}(z+y)\sigma_{3}}e^{\theta\sigma_{1}}
    e^{{1\over 2}(z-y)\sigma_{3}}\ ,\qquad
    g_C=e^{{1\over 2}(z+y)\sigma_{3}}e^{\theta\sigma_{1}}
    i\sigma_2e^{{1\over 2}(z-y)\sigma_{3}},}
where the coordinates $\theta$, $z$ and $y$ here range from
$-\infty$ to $+\infty$, and fix the gauge by setting $z=0$ (which
fixes the gauge completely since $z$ is a non-compact direction in
$SL(2,\IR)$). The background values of the metric in these two
regions were found in \ChargedBh\ to be
 \eqn\twodmetric{
    {\rm Region\ of\ type\ A:} \quad {1\over k}ds^{2} = d\theta^{2}
    -\frac{\coth^{2}(\theta)}{(\coth^{2}(\theta) -
    p^{2})^{2}}dy^{2}}
 \eqn\metAC{
     {\rm Region\ of\ type\ C:} \quad {1\over k}ds^{2} = d\theta^{2}
    -\frac{\tanh^{2}(\theta)}{(\tanh^{2}(\theta) -
    p^{2})^{2}}dy^{2}}
where
 \eqn\ppp{p^2=\tan^2\left(\frac{\psi}{2}\right) \, .}
{}For an arbitrary value of $p$ ($\psi$) one of the two regions
contains a time-like singularity line on which the $g_{yy}$
component of the metric blows up. Assuming for definiteness
$0\le\psi\le \pi/2$, that is $p^{2}\le 1$, we have a  singularity in
region C and a smooth region A. The dilaton and $U(1)$ gauge field
backgrounds read\foot{Here and below the gauge field is normalized
so that its contribution to the effective action is ${1\over
4}\int dx^2\sqrt{-g}e^{-2\Phi}F^2$.}
 \eqn\DGA{\eqalign{
    {\rm Region\ of\ type\ A:}\quad \Phi = &\Phi_{0}
    -{1\over 4}\log\({1\over 1-p^2}+\sinh^{2}(\theta)\)^2 \,
    \cr \quad A_{y} = &\frac{\sqrt{k}p}{p^{2}-\coth^{2}(\theta)} }}

 \eqn\DGC{\eqalign{{\rm Region\ of\ type\ C:} \quad \Phi =&\Phi_{0}
    -\frac{1}{4}\log\({1\over 1-p^2}-\ch^{2}(\theta)\)^2 \,
    \cr \quad A_{y} = &\frac{\sqrt{k}p}{1-p^{2}\coth^{2}(\theta)}.}}
Here $\Phi_0$ is a constant and the dilaton background value is
normalized so that $g_s=e^{\Phi}$.

%%%%%%%%%%%%%% The value of charge and mass. Connections with previous works.
The usual metric of the 2-d CBH (in Schwarzschild-like
coordinates) is obtained by the following coordinate
transformation (in regions of type A):
 \eqn\ttt{t={2y\over 1-p^2}\ ,\qquad r={2e^{-2\Phi}\over \sqrt k}~.}
In terms of $r$ and $t$ the metric and gauge field are:
 \eqn\sda{{4\over k}ds^2=-f(r)dt^2+{dr^2\over r^2f(r)}\ ,\qquad
 A_t={Q\over 2r}~,}
where
% there is a factor of $\sqrt 2$ between our gauge field
% and the one in \NP, the computation of $M$ in \NP is independent
% of the gauge field and hance holds also for our gauge field, as
% a result there is a factor $\sqrt 2$ between our Q/M and the
% one in \NP. The metric is invariant under $(M,Q,r)\rightarrow
% (aM,aQ,ar)$. {\bf (SSS)}
 \eqn\frr{f(r)=1-{2M\over r}+{Q^2\over r^2}~.}
The parameters $M$ and $Q$ are related to $p$ and $\Phi_0$ as follows
 \eqn\mmm{M={e^{-2\Phi_{0}}\over \sqrt k}\frac{1+p^{2}}{1-p^{2}}~,
 \qquad Q={2e^{- 2\Phi_{0}}\over \sqrt k}\frac{p}{1-p^{2}}~.}
The singularity is located at $r=0$.
The horizons are located at
 \eqn\rpm{r_\pm=M\pm\sqrt{M^2-Q^2}\ .}
Note that
\eqn\pprr{p^2={r_-\over r_+}~.}
The ADM mass and charge
observed from the asymptotic infinity of regions 1 and 1' are $M$ and $Q$,
while the ones observed from regions 2
and 2' are $-M$ and $Q$.

\subsec{Euclidean charged 2-d black hole background}
The Euclidean 2-d charged black hole (CBH)
is given by gauging a time-like $\widetilde{U}(1)$ in
$\widetilde{SL}(2,\IR)\times~U(1)$, where $(\sim)$ stands for the
universal cover group. The group action is
 \eqn\gaugeact{(g, x_{L}, x_{R})\mapsto
    (e^{i\rho\sigma_{2}/\sqrt{k}}ge^{i\tau\eta\sigma_{2}/\sqrt{k}},
    x_{L}+\eta \rho', x_{R} + \tau')~,}
where $\underline{\tau}=|\tau|(1,0)$, $\eta=1$ corresponds to
axial type gauging and $\eta=-1$ to the vectorial one. The anomaly
cancellation condition now implies
 \eqn\rhor{\underline{\rho} = R\underline \tau}
with
 \eqn\rmatrix{R=\(\matrix{\ch(\psi_E)& \sh(\psi_E)\cr
    \sh(\psi_E) &\ch(\psi_E)} \).}
In global coordinates, $g\in SL(2,\IR)$ is represented as follows
 \eqn\gparam{g=e^{{i\over 2}(\chi+\phi)\sigma_{2}}e^{\theta\sigma_{1}}
    e^{{i\over 2}(\chi-\phi)\sigma_{2}}~.}
To get a smooth quotient geometry we need to gauge the universal cover of
$SL(2,\IR)$ ($\widetilde{SL}(2,\IR)$), which is obtained by unwrapping
the coordinate $\chi$. We fix
the gauge by setting $\chi=0$.~\foot{If instead we fix the
gauge by setting $\phi=0$, we remain with a residual discrete
identifications as in \BBBC.} As in the Lorentzian case, we do a
KK reduction of the three dimensional coset along the compact $x$ direction.
We obtain the following two dimensional metric, dilaton and gauge field:
 \eqn\euclidgeo{\frac{1}{k}ds^{2} =d\theta^{2}+\frac{\coth^{2}(\theta)}
    {(\coth^{2}(\theta) + p^{2\eta}_{E})^{2}}d\phi^{2},}
 \eqn\phikk{\Phi=\tilde \Phi_{0}-\frac{1}{2}\log(p_E^{2\eta}\cosh^{2}
    (\theta) +\sinh^{2}(\theta)),}
 \eqn\akk{A_{\phi} ={\sqrt{k}p^\eta_{E}\over \coth^{2}(\theta) +
 p^{2\eta}_{E}}.}
%{\bf xxx (sign\ of\ A\ ?)}
The Euclidean time variable $\phi$ has a
canonical periodicity $2\pi$.~\foot{If
we gauge a finite cover of $SL(2,\IR)$, instead, we obtain an orbifold of
\euclidgeo\ with a conical singularity at $\theta=0$.}
The parameter $p_{E}^{2}$ equals
 \eqn\pe{p^{2}_{E}= \tanh^{2}\( {\psi_{E}\over 2}\)}
This geometry looks like an infinite cigar which, for $\eta=-1$,
develops a bump at $\coth^2(\theta_b)=1/p_E^2$ (see Fig. 2).
 \fig{Euclidean charged black hole.}{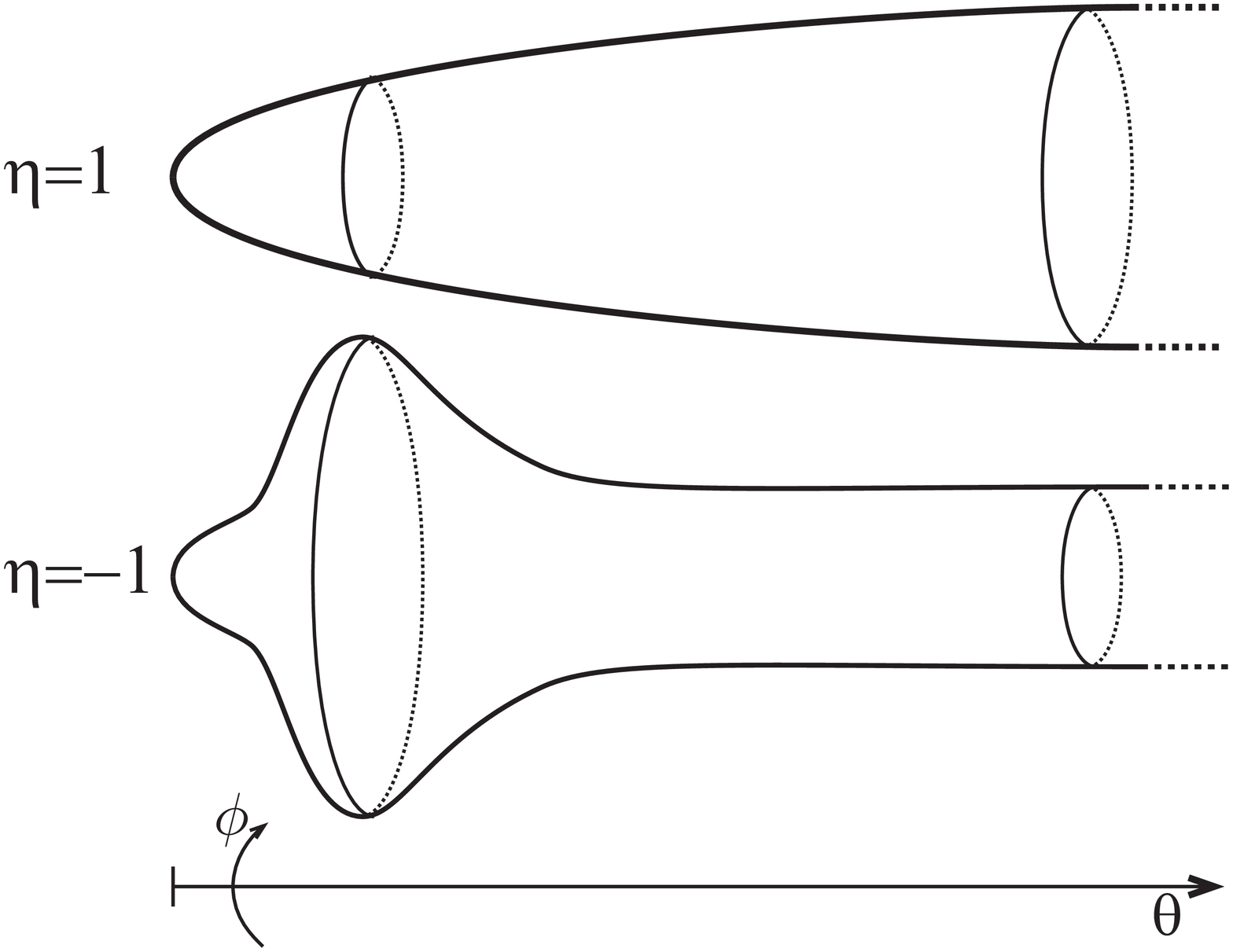}{6 truecm}
 \figlabel\cigartrampet
This Euclidean geometry \euclidgeo\ is obtained from the
Lorentzian one \twodmetric\ by the following Wick rotations:
 \eqn\WickAC{\eqalign{{\rm Region\ A} \ \ (\eta=1):&\quad y\to i\phi\, ,
\qquad \psi \to i\psi_{E}
\qquad (p^{2}\to -p_{E}^{2})\cr {\rm Region\ C} \ \ (\eta=-1):&\quad y\to
ip_{E}^{2}\phi\, , \qquad \psi \to
    i\psi_{E}  \qquad (p^{2} \to -p_{E}^{2})}}
The Euclidean metric of region A then arises from the axial gauging
($\eta=1$) while the one of region C from the vectorial gauging
($\eta=-1$). The corresponding Euclidean CFT's are related to each
other by a T-duality that acts as
$$
    T: \psi_{E} \mapsto i\pi - \psi_{E}
$$
and exchanges an axial type gauging with a vectorial one~\foot{
The vector-axial symmetry can be manifested for example by an element 
of $O(2,2, \IZ)$ (for a
review, see \GPR) together with a certain orbifolding in the case 
$\sqrt{k}p_{E}\in \IZ$. The orbifolding is expected to be a symmetry 
of the CFT as for compact parafermions.}. 
Under this transformation~\foot{Note that 
in the analytic continuation to
Minkowski space the T-duality transformation involves in addition
to $\psi \mapsto \pi - \psi$ a rescaling of the $y$ coordinate by
$p^{2}$ which is a valid operation because $y$ is noncompact in
that case. This rescaling results from changing the gauge fixing
in \slparam\ from $z=0$ to $y=0$.} \eqn\ttt{T: p_{E}^{2}\mapsto
1/p_{E}^{2}~\qquad (r_+\leftrightarrow r_-)~.} {}From the above
Euclidean metrics we read off the asymptotic radii of the
Euclidean time direction $\phi$:
 \eqn\rac{
    R^{2}_{A}(\theta\to \infty) = \frac{k}{(1+ p_{E}^{2})^{2}} \, , \qquad
    R_{C}^{2}(\theta\to \infty) = \frac{k p_{E}^{4} }{(1+ p_{E}^{2})^{2}}\,
    .}
This suggests that the temperatures
 \eqn\TT{\eqalign{T_{A} = &\frac{1}{2\pi \sqrt{\alpha'k}}(1 - p^{2})
=\frac{1}{2\pi \sqrt{\alpha'k}}{r_+-r_-\over r_+} \cr T_{C} =
&\frac{1}{2\pi \sqrt{\alpha'k}}\frac{(1 -
p^{2})}{p^{2}}=\frac{1}{2\pi \sqrt{\alpha'k}}{r_+-r_-\over r_-}}}
should be associated with the respective static regions in the
Minkowski space-time. Here we have restored the dimensionfull unit
-- the string length scale $\sqrt{\alpha'}$ -- using the fact that
$R_{AdS_{3}}/\sqrt{\alpha'}=\sqrt{k}$.

When we take the uncharged limit $p,\ p_E \to 0$ (equivalently,
when the inner horizon coincides with the singularity $r_-\to 0$),
the Euclidean metric outside the black hole event horizon becomes
(see eq. \euclidgeo\ with $\eta=1$):
 \eqn\uncharged{{1\over k}ds^{2}= d\theta^{2} +\tanh^{2}(\theta)
 d\phi^{2}~.}
The temperature of the uncharged black hole, as read from
\uncharged, \TT, is $T_A=(2\pi\sqrt{\alpha' k})^{-1}$. On the
other hand, the $p,\ p_E \to 0$ limit of the region beyond the
singularity is degenerate (see eq. \euclidgeo\ with $\eta=-1$):
\eqn\bslim{{1\over k}ds^{2}\to d\theta^{2} +
p_E^4\coth^2(\theta)d\phi^2~, \qquad p_E\to 0~.} This background
is ill defined for two reasons. First, formally, the radius of the
Euclidean time direction $\phi$ vanishes as $p_E$ approaches zero
(which implies, in particular, $T_C\to\infty$). Second, there is a
curvature singularity at $\theta=0$. Hence, the temperature $T_C$
associated with the region beyond the singularity is ill defined
in the uncharged limit -- there is no horizon and no
straightforward way to define a smooth Euclidean continuation of
the region beyond the singularity of the uncharged black
hole\foot{Note that when $p_E=0$ and $\eta=-1$, $\chi$ cannot be
changed by a gauge transformation and hence $\chi=0$ cannot be
fixed in \gparam. Instead, one can fix the gauge by setting
$\phi=0$. The metric one gets is: ${1\over
k}ds^{2}= d\theta^{2} + \coth^{2}(\theta) d\chi^{2}$. If, for
instance, we gauge $SL(2,\IR)$ instead of $\widetilde{SL}(2,\IR)$
then $\chi$ is compact with periodicity $2\pi$, and this
background looks like a trumpet \refs{\giv,\dvv}
which degenerates at $\theta=0$
and with asymptotic radius $R_{\infty}/\sqrt{\alpha'} = \sqrt{k}$.
But in any case, since there is no tip and hence no conical
defect, it is not well defined a priori which is the appropriate
Euclidean background obtained by the Wick rotation of the region
beyond the singularity of the uncharged black hole
\refs{\Elitzur,\witten,\giv,\dvv,\GPR}.}. This result is perhaps
compatible with the expectation \refs{\dvv,\ChargedBh,\gkos} that
in the uncharged case the singularity (as seen from the
asymptotically flat region behind it) is (classically) a perfect
reflector and hence the low energy physics beyond the singularity
cannot communicate with any other region; in particular, this
suggests that there is no entanglement between regions 2 and 2'.

The emergence of a compact dimension in the presence of a gauge
potential suggests that a thermodynamical characteristic could be
attributed to the value of a Wilson loop around the compact
direction (see \BBW\ for different view on the same point). In
search for such a meaning one notes that another thermodynamical
characteristic of the CBH is its electric chemical potential.
Adding a charge $q$ point-like particle at rest in the black hole
frame at $\theta_0$ corresponds to adding a charge density\foot{In
general for a point-like particle $J^\mu({\bf x})=q\int{{\d
y^\mu\over\d \tau}{\delta^d\({\bf x-y}(\tau)\)\over
\sqrt{-g}}d\tau}$, where $\tau$ is its proper time. The
corresponding contribution to the action is $\delta I=-\int
dx^d\sqrt{-g}J^\mu A_\mu=-q\int{\d y^\mu\over\d \tau}A_\mu d\tau$,
which for a particle at rest equals $-q\oint dy^0
A_0(\vec{r}_0)$.}
$$
    J^{0}=q{\delta(\theta-\theta_0)\over\sqrt{-g}}\ , \qquad J^i=0~,
$$
which contributes to the Euclidean action
%{\bf xxx (to check the sign)}
 \eqn\wil{-\int d^2x\sqrt{-g}J^\mu A_\mu=-q\oint d\phi
 A_\phi(\theta_0)~.}
If (following \refs{\Gibbons,\GP,\NP}) we identify the value
of the action on the equations of motion with the thermodynamical
free energy
$$
    -I_{|_{\rm{on\ shell}}}=\beta F=\beta M-S-\beta\sum_i\mu_i{\cal Q}_i~,
$$
then the Wilson loop \wil\ at infinity is given by
 \eqn\www{W=\exp\(\oint d\phi A_\phi(\theta\to \infty)\)=\exp
    (\beta\mu_{\rm{el}})~.}
In the Lorentzian geometry the value of the gauge field at
infinity could have assumed ab initio any value; different values
are related by large gauge transformations. In the Euclidean
geometry the angular part of the vector field $A_\mu$ (which is
$A_\phi$) has to vanish at the tip of the cigar $\theta=0$. This
Euclidean regularity condition at the tip dictates a particular
non-vanishing constant value of $A_\phi$ at infinity. The value
obtained is the same one obtained from the Euclidean coset,
 \eqn\Aac{
    A_\phi^{(A)}(\theta\to \infty)=A_\phi^{(C)}(\theta\to
    \infty)={\sqrt k p_E\over p_E^{2}+1}~,}
which leads to the following values of the Wilson loops:
 \eqn\wilval{W_A=W_C=\exp\(\frac{2\pi\sqrt{k}p_E}{p_E^{2}+1}\)~.}
Combining \TT\ and \wilval\ one reads the corresponding chemical
potentials:
 \eqn\chem{
    \mu_A=-p=-{Q\over r_+}\ , \qquad
    \mu_C={1\over p}={Q\over r_-}~,}
which are in agreement with \NP. In the Lorentzian
space this electro-chemical potential is the electric potential on the
horizon of the black hole (outer and inner, respectively).

One can repeat the computation for a RN black hole in an arbitrary
dimension $d\ge 4$. When analytically continuing a region outside
of the black hole to a compact Euclidean geometry the horizon
becomes the tip of the $d$-dimensional cigar\foot{The Euclidean
geometry is a cigar-like in the radial and Euclidean time
coordinates times a $(d-2)$-dimensional sphere which shrinks to
zero size at the tip.}. Naively, the value of $A_0$ at the tip is
the electric potential at the horizon, but since $A_\mu$ is a
vector field its angular component has to vanish at the tip. We
learn that in order to have a non-singular gauge field at the
Euclidean tip, in the Lorenzian black hole we have to do a large
gauge transformation which gives a vanishing gauge field at the
horizon (which can be done in the Lorenzian case). After the large
gauge transformation the gauge field will not vanish at infinity
anymore (for $d\ge 4$), but instead its value at infinity is minus
the electric potential at the horizon. After analytic continuation
to the Euclidean cigar this asymptotic value of the gauge field at
infinity leads to a non-trivial Wilson loop. The value of the
Wilson loop is minus the electric potential at the horizon times
the circumference of the compact time, which is the inverse black
hole temperature. This computation gives the following electro-chemical 
potential for an RN black hole in dimension $d\ge 4$:
 \eqn\chimd{\mu_{\rm el}=-{Q\over r_+^{d-3}}~.}

Requiring a non-singular metric for a Euclidean black hole led to
a compact Euclidean dimension. Requiring a non-singular gauge
field (at the tip) led to a non-trivial Wilson loop at infinity.
This compact dimension and a non-trivial Wilson loop both reflect
the surprising feature of black holes -- they behave as a
thermodynamical system. The temperature associated with them is
encoded in the asymptotic radius of the compact direction and the
electro-chemical potential is encoded in the asymptotic Wilson
loop wrapped around the compact direction. In a similar way one
can compute the chemical potential for rotating particles in a Kerr
black hole; this is demonstrated in appendix A.

Back to 2-d, another thermodynamics characteristic of the 2-d CBH
is its entropy. By integrating the equations
 \eqn\temp{{1\over T}=\({\d S\over\d E}\)_Q \ ,\qquad \mu=-T\({\d S\over\d
    Q}\)_M\ ,}
where $E_A=M^{(A)}_{\rm ADM}=M$ and $E_C=M^{(C)}_{\rm ADM}=-M$, we
find
 \eqn\entrop{
    S_A=\pi \sqrt{k}\ r_+\ ,\qquad
    S_C=\pi \sqrt{k}\ r_-~.}
The heat capacities in regions A and C are read from the relation
$C=\(\d E\over\d T\)_Q=T\({\d S\over \d T}\)_Q$:
 \eqn\hetcapaci{\eqalign{
    C_A=&{\pi\sqrt k (r_+-r_-)\over 2}{r_+\over r_-}\cr
    C_C=&-{\pi\sqrt k (r_+-r_-)\over 2}{r_-\over r_+}~.}}
Note that $S_C\neq S_A$ (except for the extremal case $M=|Q|$
($r_+=r_-$)), and $C_C\neq C_A$. Moreover, $C_A$ is positive while
$C_C$ is negative. The quantities $S_C$ and $C_C$ are, however, in
agreement with T-duality \ttt, which interchanges
$r_+\leftrightarrow r_-$.~\foot{Or, equivalently, $M\to -M$
(recall that $r_{\pm}=|M\pm \sqrt{M^2-Q^2}|$).}

To summarize, low energy observers beyond the singularity see a negative mass
object with a naked singularity, with a negative specific heat and
an entropy which is lower than the black hole entropy as seen by
an observer outside the event horizon. The interpretation of these
results is not clear to us, though it is possible that string
theory manages to ``take care of itself'' even in such a
situation; the agreement between the thermodynamical and the
T-duality derivations motivates this logical
possibility~\foot{T-duality implies that while the correlators of
momentum modes on the Euclidean background corresponding to region
A (\euclidgeo\ with $\eta=1$) describe, via analytic continuation,
the physics of momentum modes outside the event horizon of the
Lorentzian background, the correlators of winding modes in A
describe instead, via analytic continuation, the physics of
momentum modes in the region beyond the inner horizon.
Alternatively, the latter is described by analytic continuation of
momentum modes on the T-dual Euclidean background corresponding to
region C (\euclidgeo\ with $\eta=-1$). The mysterious
thermodynamical quantities, in the region with $\eta=-1$, are obtained
from this T-dual Euclidean description.}.

Note that the T-duality we have been exploiting when comparing the
thermodynamical quantities is a duality between Euclidean
backgrounds. Thus, a priori it does not have to be a true symmetry
of the Minkowski signature background\foot{One could use T-duality
along the time-like direction of the Minkowski signature
background. This T-duality however is known to suffer, in some
cases, from various kinds of problems, see e.g.~\Hull.}. However,
even if we assume it is a symmetry (not relying on a would be
thermodynamics of negative mass naked singularities), it is not obvious
that the total entropy of the two Euclidean systems measures the
same thing. If nevertheless the entropy turns out to be T-duality
invariant it should be expressed in terms of a symmetric function
in $S_{A}$ and $S_{C}$, such as for instance their sum. This would
imply that there is more entropy to the system than seen above in
a single region. A different possibility, argued in \VW, is that
the total entropy of the system is the (absolute value of the)
difference (this result assumed that  degrees of freedom do not
reside beyond the inner horizon).

%%%%%%%%%%%%%%%%%%%%%%%%%%%%%%%%%%%%%%%%%%%%%%%%%%%%%%%%%%%%%%%%%%
%                 MINKOWSKI    COMPUTATION
%%%%%%%%%%%%%%%%%%%%%%%%%%%%%%%%%%%%%%%%%%%%%%%%%%%%%%%%%%%%%%%%%%

\subsec{Minkowski signature computation of the temperatures}

For the model at hand we know explicitly the scalar field
(tachyon) wave functions in all space-time regions including the
type C static region with singularity~\foot{At the moment, this is
not the case with the 4-d Reissner-Nordstrom solution.}. The wave
functions for 2-d CBH as well as a scattering problem set in the
region behind the singularity were recently discussed in
\ChargedBh. The wave functions essentially descend from the
Laplace eigenfunctions on the $SL(2, \IR)$ group manifold. In
particular, they are smooth on the singularity lines \refs{\ChargedBh,\gkos}. 
One can therefore perform an alternative computation of the
temperatures in regions A and C via Bogolubov transformations
relating the analogs of Rindler and Minkowski vacua on outer and
inner horizons, respectively. A computation of this kind was done
by Israel in \Israel\ for eternal 4-d black holes. The presence of
charge does not alter the computation -- the essential feature
being the fact that the Killing vector is globally time-like in a
static region on each side of the horizon (see e.g. \Waldbook). 
If one considers charged particles on the background
at hand the corresponding electro-chemical potential \chem\ will enter the
canonical distribution. Here we are interested mainly in the
uncharged field case characterized by an Hawking temperature only.
The arguments of \Israel\ are then directly applicable to region A
or region C separately.

%%%%%%%%%%%%%%%%%%%%%%%%%%%%%%%%%%%%%%%%%%%%%%%%%%%%%%%%%%%%%%%%%%
Let us outline here the ingredients of the computation in \Israel.
We consider a static region of space-time that consists of two
patches: ${\cal U}^{-}$, ${\cal U}^{+}$ -- the left and the right
wedges separated by a bifurcate Killing horizon. In two dimensions
the last one is comprised out of two intersecting light rays ${\bf
h}^{+}$, ${\bf h}^{-}$ (see Fig. 3).
 \fig{Bifurcate Killing horizon.}{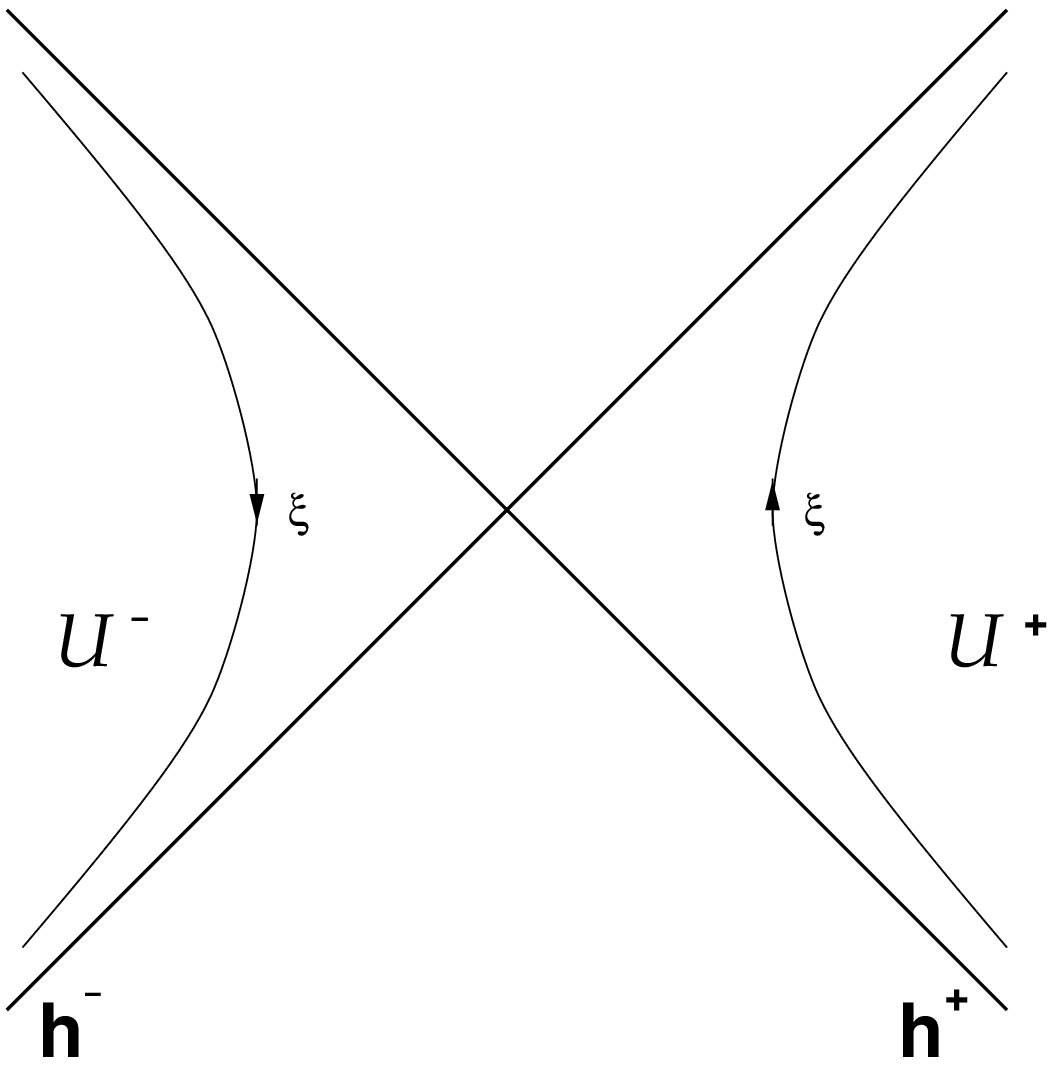}{6 truecm}
 \figlabel\horizon
The Killing vector $\xi$ in ${\cal U}^{-}$ and ${\cal U}^{+}$ is
assumed to be everywhere time-like. One considers then two
quantizations in ${\cal U}^{-}\cup {\cal U}^{+}$. One is built
upon ``Killing modes" which are defined as energy eigenfunctions of
the Klein-Gordon equation as registered by a Killing orbit
observer in each patch ${\cal U}^{\pm}$. The corresponding
one-particle Hilbert spaces ${\cal H}^{-}$, ${\cal H}^{+}$ consist
of solutions which are purely negative frequency with respect to
Killing time translations\foot{The positive/negative frequencies
in region ${\cal U}^{-}$ are defined w.r.t $-\xi$.}. Thus the
second quantized Fock space defined by Killing time modes can be
written as a direct product
$$
    {\cal F}({\cal H}^{-})\otimes {\cal F}({\cal H}^{+}) \, .
$$
The corresponding vacuum $|0\rangle_{-} \otimes |0\rangle_{+}$ is
an analog of the Boulware vacuum of an extended Schwarzschild
solution. In the other quantization an analog of the
Hartle-Hawking vacuum $|HH\rangle $ is defined via Klein-Gordon
eigenfunctions whose restrictions on the light rays ${\bf h}^{+}$,
${\bf h}^{-}$, comprising the horizon, contain positive frequency
modes with respect to the affine parameters along the rays. It
follows from a standard computation \Israel\ of Bogolubov
coefficients relating the two quantizations that the
Hartle-Hawking vacuum can be represented as an entangled state
 \eqn\HH{
    |HH\rangle = 
\prod_{\omega,q}\sum_{n=0}^{\infty}\exp\(-{n\pi \over \kappa}
(\omega-\mu_{\rm el}q)\)
    |n, \omega,q\rangle_{-}\otimes |n, \omega,q\rangle_{+} \in
    {\cal F}({\cal H}^{-})\otimes {\cal F}({\cal H}^{+})~,}
where $|n, \omega,q\rangle_{\pm}$ stands for a state with $n$
``Killing particles" with energy $\omega$ and charge $q$. Here
$\kappa$ is the surface gravity constant characterizing the
Killing horizon at hand and $\mu_{\rm el}$ is the electric potential (at
the horizon). Restriction of the state \HH\ to a single patch
${\cal U}^{+}$ (or ${\cal U}^{-}$) yields a thermal density matrix
characterized by an Hawking temperature
 \eqn\sgr{T=\frac{\kappa}{2\pi} \, .}
The main assumption in the computation leading to \sgr\ is that
the horizon is a non-singular part of space-time. The analog of
Hartle-Hawking vacuum \HH\ can be characterized by the requirement
of its invariance under the isometries generated by the Killing
vector at hand and by the regularity of the scalar field
stress-energy tensor on the horizon (see e.g. \Waldbook\ for a
detailed discussion).

The above considerations, being applied separately to region $A$ or
$C$ of the 2-d CBH solution, yield Hawking temperatures
$T_{A,C}=\frac{\kappa_{_{A,C}}}{2\pi}$, where $\kappa_{_{A,C}}$ is
the surface gravity of the outer (for region A) or inner (for
region C) horizon. Noting that the Killing vector, normalized to
have norm one at infinity, is in both regions
 \eqn\killing{\xi \equiv \pm
    {(1-p^{2})\over\sqrt k} \frac{\partial}{\partial y}~,}
one can read off the surface gravity from the asymptotic expansion
of the metric in the vicinity of the corresponding horizons:
 \eqn\asymmet{\eqalign{{\rm Region\ of\ type\ A:} \quad &\frac{1}{k}
ds^{2} \approx d\theta^{2} -\theta^{2}dy^{2}\cr {\rm Region\ of\
type\ C:} &\quad \frac{1}{k}ds^{2} \approx d\theta^{2}
-\frac{\theta^{2}dy^{2}}{
 p^{4}} \, .}}
One obtains then the same result as in \TT, as expected.

%Both temperatures can
%be written in terms of positions of the horizons in the
%Schwarzschild type coordinates.
% \eqn\tac{
%    T_{A} = \frac{1}{2\pi \sqrt{\alpha'k}}\frac{(r_{+}-r_{-})}{r_{+}} \, ,
%\qquad
%    T_{C} = \frac{1}{2\pi \sqrt{\alpha'k}}\frac{(r_{+}-r_{-})}{r_{-}} \,.}
%This representation makes it manifest that the temperatures are
%properties of the horizons.

%%%%%%%%%%%%%%%%%%%%%%%%%%%%%%%%%%%%%%%%%%%%%%%%%%%%%%%%%%   LOCAL TEMPERATURES

The temperatures considered above correspond to the Hawking
radiation temperatures as measured
by  asymptotic observers in region A and C. In addition to that,
one can consider temperatures measured locally by observers moving
along  Killing vector orbits. Those temperatures are obtained by
dividing the asymptotic temperatures by the Killing vector norm.
We then have
 \eqn\ta{T_{A}^{local}=\frac{|\coth^{2}(\theta)-p^{2}|}{
2\pi\sqrt{\alpha'k}\coth(\theta)} \, ,}
 \eqn\tc{T_{C}^{local}=\frac{|\tanh^{2}(\theta)-
 p^{2}|}{2\pi\sqrt{\alpha'k}p^{2}\tanh(\theta) } \, .}
We see that both temperatures characteristically go to infinity on
the corresponding horizons. In addition, in region C the local
temperature vanishes on the singularity line.

%%%%%%%%%%%%%%%%%%%%
In this section we have followed the point of view of \Israel\ in
which the boundary conditions were set up on the horizon. Note
that a situation with a clear separation between the Hilbert space
two components is considered in \Maldacena. A pair of 
Hilbert spaces of regions 1 and 1' will be entangled with the
Boltzman weight appropriate for the temperature $T_A$ \TT. Since
regions 1 and 1' are outside the event horizons, while regions 2
and 2' lay beyond the inner horizons, a common wisdom would be that
the latter regions are not involved in the former entanglement. An
alternative entangled state could be built out of states belonging
to the Hilbert spaces of regions 2 and 2' and weighted in that
case by the Boltzman factor corresponding to the temperature $T_C$
\TT. Such states may well reproduce correlation functions between
operators evaluated either only in regions 1 and 1' or only in
regions 2 and 2'. On the other hand, to reproduce mixed
correlators such as one between an operator inserted in region 1
and another inserted in region 2, one has to quantize the field in
all regions. An attempt to do this is the subject of the next
section.

%%%%%%%%%%%%%%%%%%%%%%%%%%%%%%%%%%%%%%%%%%%%%%%%%%%%%%%%%%%%%%%%%%%%%%
%                      Q U A N T I Z A T I O N
%%%%%%%%%%%%%%%%%%%%%%%%%%%%%%%%%%%%%%%%%%%%%%%%%%%%%%%%%%%%%%%%%%%%%%

\newsec{Quantization of a scalar field in the extended charged BH
 background}
The discussion in the previous section concerned with the static
patches $1\cup 1'$ and $2\cup 2'$, each one considered separately
from all other regions in the extended Penrose diagram of
Fig. \cbh. In this section we consider a scalar field quantization
in the space-time region that includes all together regions
$1,1',2,2'$ and II of Fig. \cbh. These regions, together with
region III, represent a basic block of the maximally
extended space-time (that includes infinitely many blocks of the
same kind). The space-time region at hand is not globally
hyperbolic and is not everywhere static\foot{Quantization in
static non-hyperbolic space-times was recently discussed in a
series of papers \Wald.
The rigorous results obtained in those papers do not apply
however to our situation.}.
We thus have to deal with a rather
non-standard situation here. We will present two possible
approaches to quantization on this space-time. The first one
follows more closely the common wisdom of canonical quantization
in curved space-time, while the second one relies on the $SL(2,
\IR)$ structure underlying the model at hand and is more string
theory inspired.

\subsec{Quantization 1} Let us start by outlining the main steps
in which we will proceed in this subsection. First we discuss
general features of the quantization on the extended space-time at
hand: the initial data surface that sets up the first quantized
Hilbert space, the canonical commutation relations and the way the
commutation relations for the field modes can be determined. A
state, analogous to the Hartle-Hawking vacuum, can be specified by
requiring its invariance under the isometry (generated by $\xi$
\killing) and the regularity of the stress-energy tensor
expectation value (related to the Hadamard condition \Wald ).
Correlators in such a state will be thermal with temperatures
$T_{A}$ and $T_{C}$ for operators inserted only in the region $A$
or $C$, respectively. Such a state, if it exists, has to be
annihilated by certain operator modes. We construct two such
operator modes whose commutator, computed with the help of
canonical commutator relations, is a non-vanishing $c$ number. This
will lead us to the conclusion that a state that would be
analogous, in the sense described above, to the Hartle-Hawking
vacuum for the whole extended space-time at hand does not exist.

 \fig{Initial data surface for the extended RN space-time.}
 {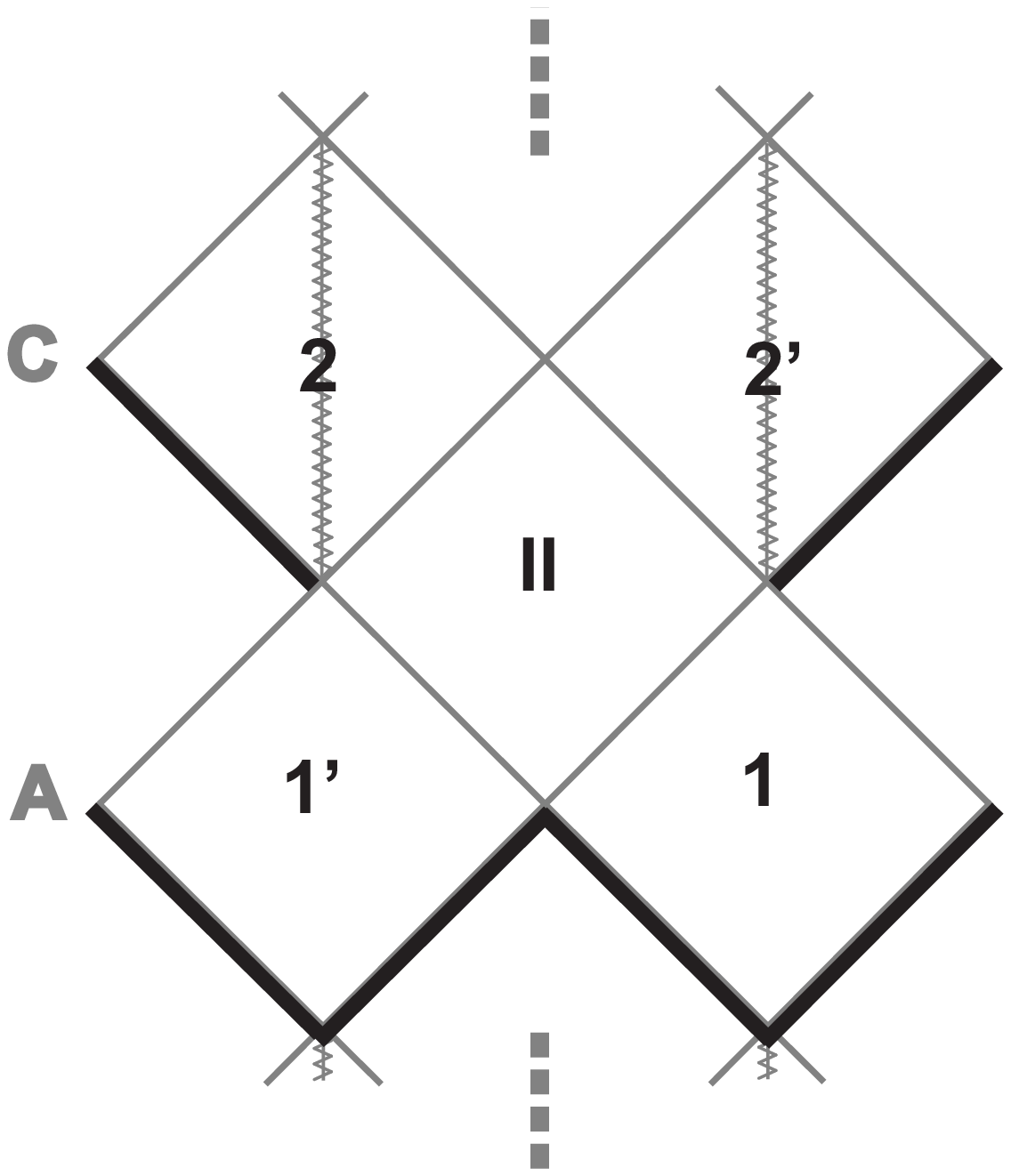}{6 truecm} \figlabel\cauchy
To define the first quantized Hilbert space of wave functions one
can choose to specify the initial data for the Klein-Gordon
equation on the asymptotic null-surfaces that have no causal past.
This singles out the null-lines boldfaced in Fig. \cauchy. We will
assume that the corresponding null initial value problem is
mathematically well defined\foot{One of the concerns on the
mathematical side is what condition substitutes the usual
smoothness assumption of the initial data when working with
disconnected initial data surfaces as in the case at hand. Some
questions related to the initial value problem in the region
containing a singularity were  considered in \gkos, where in
particular it was shown that for the initial data specified on the
past null infinities  there is a unique solution to the wave
equation defined on both sides of the singularity.}. We are thus
exploring the possibility that a general pure state in this
extended Hilbert space is of the form
 \eqn\generalstate{|\Psi\rangle=\sum_{ijkl}C_{ijkl}|i\rangle_2
 |j\rangle_{1'} |k\rangle_1 |l\rangle_{2'}~,}
where, for each $\lambda\in \{1,1',2,2'\}$, the states
$|j\rangle_{\lambda}$ form a complete basis for wave functions
supported on the part of the initial data surface lying in
%(or on the boundary of)
region $\lambda$ and vanishing on the rest of the
initial data surface (see Fig. \cauchy).

Let $f_{\omega,l}$ be a complete basis to the space of solutions
of the massless Klein-Gordon equation in the extended region of
space-time at hand (with their initial data specified on the
disconnected Cauchy surface of Fig. \cauchy). The functions in
this basis are labelled by the Killing vector eigenvalue $\omega$,
and an additional parameter $l$ that takes finitely many values.
In our discussion we are not going to use any specific basis. In
appendix B we explain how to construct a basis of wave functions
which are analytic with respect to the affine parameters on the
horizons.

A quantum field in the extended space-time can be expanded as
\eqn\ph{
 \phi(x) = \sum_{l}\int_{0}^{\infty}d\omega \, a_{\omega,l} f_{\omega, l}
+ \enspace {\rm h.c.}~,
}
where $a_{\omega,l}$ are operators whose commutation relations
are to be determined. The canonically conjugated momentum is an
operator $\Pi(x,t)=\sqrt{-g}g^{tt}\partial_{t}\phi(x,t)$ that
should satisfy the canonical equal time commutation relation
 \eqn\CCR{[\phi(x_{1},t), \Pi(x_{2},t)] = i\delta(x_{1}-x_{2}) \,.}
Also, we require that the standard causality assumption holds by
which a commutator of the quantum field $\phi(x)$ with itself
vanishes for space-like separated points.

\fig{The fixed-time slices $\Sigma_{A}$, $\Sigma_{C}$ and
 $\Sigma_{II}$.}{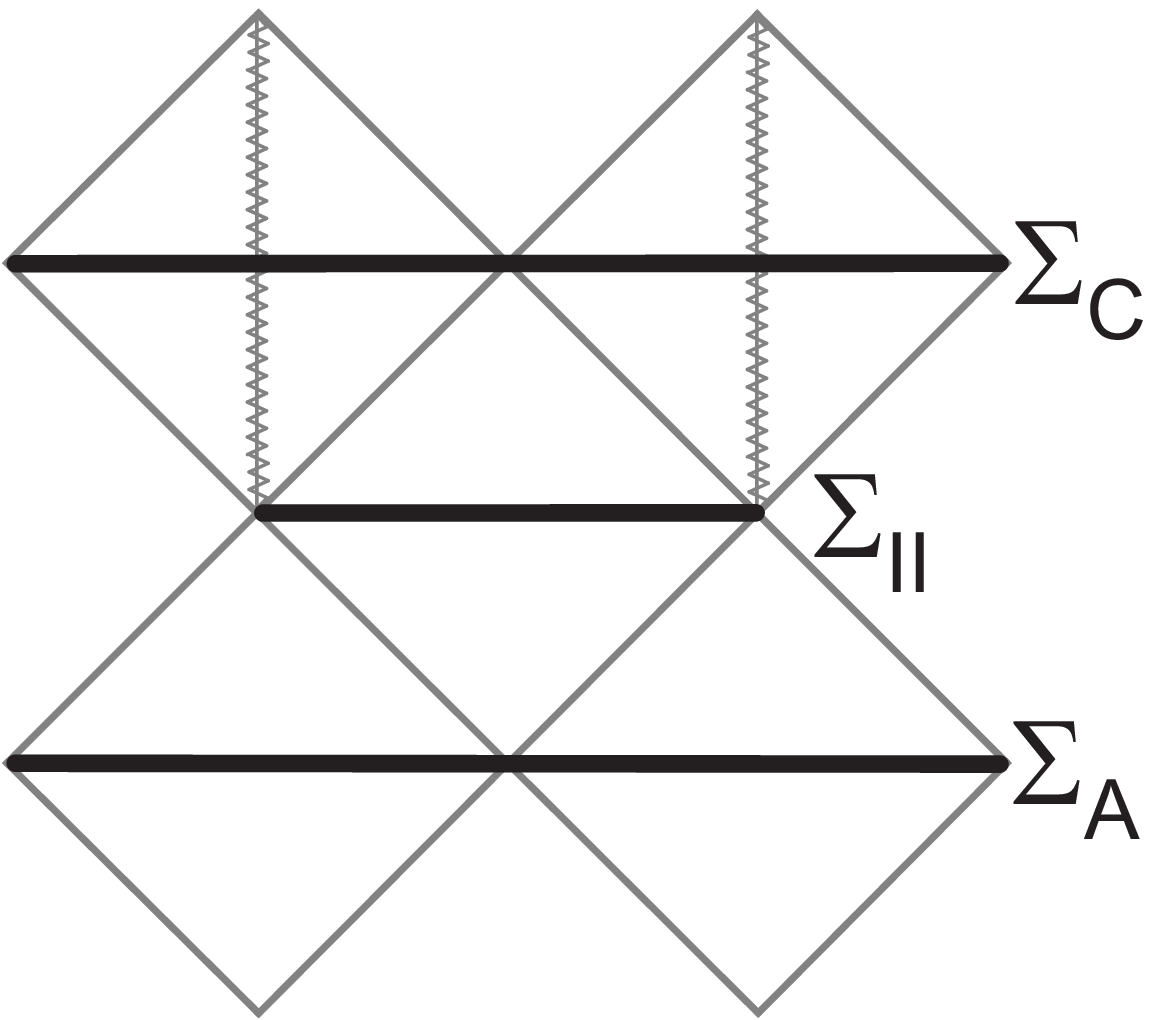}{6 truecm}
 \figlabel\KG
Let us further introduce three fixed-time slices $\Sigma_{A}$,
$\Sigma_{C}$, $\Sigma_{II}$ depicted in Fig. \KG. Denote
 \eqn\KGn{ (f, g)_{\Sigma_{\delta}} = i\int_{\Sigma_{\delta}}dx
g^{tt}\sqrt{-g} ( \bar f(x,t)\partial_{t}g(x,t)  -
 g(x,t)\partial_{t}\bar f(x,t) ) }
-- the Klein-Gordon inner products evaluated on the corresponding
slices $\delta\in \{A,II,C\}$. By canonical commutation relations
\CCR\ and local causality we have on each slice $\Sigma_{\delta}$:
 \eqn\nicef{ [(\phi, f)_{\Sigma_{\delta}},
 (\phi,\bar{g})_{\Sigma_{\delta}}] = (g,f)_{\Sigma_{\delta}} \, .}
Noting that a Killing vector eigenmode is fixed by its eigenvalue
$\omega$ and its value on the slices $\Sigma_A$ and $\Sigma_C$, we
see that substituting the expansion \ph\ into \nicef\ for each
slice $\Sigma_{\delta}$ allows one, in principle, to express
commutation relations between all of the mode operators
$a_{\omega,l},\ a_{\omega,l}^\dagger$ in terms of the inner products
$(f_{\omega,l}, f_{\omega',l'})_{\Sigma_{A}}$ and $(f_{\omega,l},
f_{\omega',l'})_{\Sigma_{C}}$, whose explicit form depends on the
choice of basis at hand.

In our discussion we rely on theorems proved by Kay and Wald
\KayWald\ regarding the uniqueness of the Hartle-Hawking state in
a globally hyperbolic static space-time with bifurcated horizons.
In the course of the proof an algebraic state is defined as a
mapping from an algebra of observables to complex numbers,
satisfying the necessary properties of an average value (see
\Waldbook\ for details). In particular, such a definition of a
state includes pure and mixed states realized in the standard
approach to second quantization via the Fock space construction.
The results of Kay and Wald apply to globally hyperbolic static
regions with a (single) bifurcated Killing horizon. It is shown in
\KayWald\ that if an algebraic state is invariant under the
isometry (generated by a time-like Killing vector) and if the
expectation value of the stress-energy tensor in this state is
regular on the horizon, then this state is a pure state and it is
unique. If furthermore there is a discrete isometry relating the
static patches on the two sides of the horizon, a restriction of
the state to one of the patches is thermal. It is called the
Hartle-Hawking state; we denote the corresponding Hartle-Hawking
state for regions A and C by $|\widetilde{HH}\rangle_A$ and
$|\widetilde{HH}\rangle_C$, respectively.
%In the case of quantum field theory in globally hyperbolic region
%ith a single horizon (regions of type A (B) and their causal
%uture) a unique algebric state, if it exist, which is invariant
%under the isometry (generated by $\xi$) and in which the expected
%stress-energy tensor is regular on the horizons (Hadamard
%condition \Wald) is the Hartle-Hawking state
%$|\widetilde{HH}\rangle_A$ ($|\widetilde{HH}\rangle_B$) \KayWald,
%which is pure.
 Given the space of classical solutions with initial data on a
Cauchy surface and a symplectic structure one can define a field
theory Hilbert space via specifying  positive and negative
frequency wave functions  \Waldbook. The Fock vacuum is a state
which is annihilated by all the positive frequency modes. The
Hartle-Hawking (pure) state is the Fock vacuum where positive and
negative frequencies are defined with respect to the affine
parameters along the (single) bifurcate Killing horizon. Here we
describe an attempt to find  an analog to the Hartle-Hawking state
$|\widetilde{HH}\rangle$ for the extended CBH background. We start
from a general pure  state\foot{This assumption is made only for
the sake of simplifying the argument; it can be also shown to hold
for a mixed state.} in an extended Hilbert space
 \eqn\hh{|\widetilde{HH}\rangle=\prod_{i}\(
 \sum_{n_{i}^{(2)}\!\!, n_{i}^{(1')}\!\!, n_{i}^{(1)}\!\!, n_{i}^{(2')}}
 C(n_{i}^{(2)}\!\!, n_{i}^{(1')}\!\!, n_{i}^{(1)}\!\!, n_{i}^{(2')})
 |n_{i}^{(2)}\rangle_2  |n_{i}^{(1')}\rangle_{1'}
 |n_{i}^{(1)}\rangle_1 |n_{i}^{(2')}\rangle_{2'}\)}
where $|n_{i}^{(2)}\rangle_2  |n_{i}^{(1')}\rangle_{1'}
|n_{i}^{(1)}\rangle_1 |n_{i}^{(2')}\rangle_{2'}$ stands for a
state in the Fock space with $n_{i}^{(\lambda)}\in \IN$ excited
modes $|i\rangle_{\lambda}$. Restricting the state
$|\widetilde{HH}\rangle$ to the regions of type A (C) of Fig.
\cauchy\  one obtains an algebraic state in these regions. We mark
this {\it algebraic} state by $\widetilde{HH}_A$
($\widetilde{HH}_C$). If $|\widetilde{HH}\rangle$ is invariant
under the isometry and its expected stress-energy tensor is
regular on the horizons, so is $\widetilde{HH}_A$
($\widetilde{HH}_C$). Note that regions A and C are static and
globally hyperbolic (for the class of initial conditions  in
region C obtained by setting up the data on the past null
infinities \gkos). Thus, according to \KayWald, $\widetilde{HH}_A$
($\widetilde{HH}_C$) have to be equal to
$|\widetilde{HH}\rangle_A$ ($|\widetilde{HH}\rangle_C$). Strictly
speaking the theorems of Kay and Wald may not be applicable for
the region C which has singularities. However, as our first concern
is the behavior of the stress-energy tensor on the horizons we may
restrict our considerations to a globally hyperbolic subregion in
between the singularities that contains the bifurcation point of
the horizon. In this subregion the results of Kay and Wald are
directly applicable. This means that the restriction of
$|\widetilde{HH}\rangle$ to such a subregion $\widetilde C$ will
be equal to the corresponding Hartle-Hawking state
$|\widetilde{HH}\rangle_{\widetilde C}$. In what follows we will
not restrict to such a subregion although all our arguments apply
for such a restriction.

We see that if we consider a restriction of the field $\phi(x)$ to
region $A$ its positive frequency components $(\phi,
g_{\omega})_{\Sigma_{A}}$, defined with respect to affine times on
the outer horizon, have to annihilate $|\widetilde{HH}\rangle$.
Here $g_{\omega}$ is an arbitrary Killing vector eigenmode that
contains only positive frequency modes with respect to affine
times on the outer horizon. Analogously, the operators $(\phi,
h_{\omega})_{\Sigma_{C}}$ have to annihilate
$|\widetilde{HH}\rangle$ for any function $h_{\omega}$ comprised
out of positive frequencies which are defined only with respect to
affine times on the inner horizon. We are going to demonstrate
that these conditions are not mutually compatible by constructing
two such annihilation operators that commute to a $c$ number. To
that end let us construct a specific mode $F_{\omega}(x)$ that has
the property that it consists of pure positive affine time
frequencies in region $A$ and of pure negative affine time
frequencies in region $C$ (the affine times being defined on the
outer and inner horizons, respectively).

The state $|\widetilde{HH}\rangle$ then has to be annihilated by
operators $(\phi, F_{\omega})_{\Sigma_{A}}$, $(\phi, \bar
F_{\omega})_{\Sigma_{C}}$ where $\bar F_{\omega}$ stands for the
complex conjugate function and hence contains only positive
frequency modes with respect to affine times on the inner horizon.
The commutator of these two operators can be computed as follows.

First note that by Klein-Gordon norm conservation we have
\eqn\KGcons{\eqalign{
 (\phi, F_{\omega})_{\Sigma_{A}}
&= (\phi, F_{\omega})_{{\cal I}_{1}^{+}} +(\phi,
F_{\omega})_{\Sigma_{II}} + (\phi, F_{\omega})_{{\cal I}_{1'}^{+}}
\cr (\phi, \bar F_{\omega})_{\Sigma_{C}} &=  (\phi, \bar
F_{\omega})_{{\cal I}_{2}^{-}} + (\phi, \bar
F_{\omega})_{\Sigma_{II}} + (\phi,
 \bar F_{\omega})_{{\cal I}_{2'}^{-}}  }}
and analogous expressions for other operators. Here ${\cal
I}_{1}^{+}$, ${\cal I}_{1'}^{+}$ stand for asymptotic future
null-infinities of regions $1$ and $1'$, respectively, and ${\cal
I}_{2}^{-}$, ${\cal I}_{2'}^{-}$ denote the asymptotic past
null-infinities of regions $2$ and $2'$; see
Fig. 6. %Fig. \scray.
\fig{The null surfaces ${\cal I}_{1}^{+}$, ${\cal I}_{1'}^{+}$
  stand for asymptotic future null-infinities of regions $1$ and
  $1'$, respectively, and ${\cal I}_{2}^{-}$, ${\cal I}_{2'}^{-}$
  denote the asymptotic past null-infinities of regions $2$ and
  $2'$.}{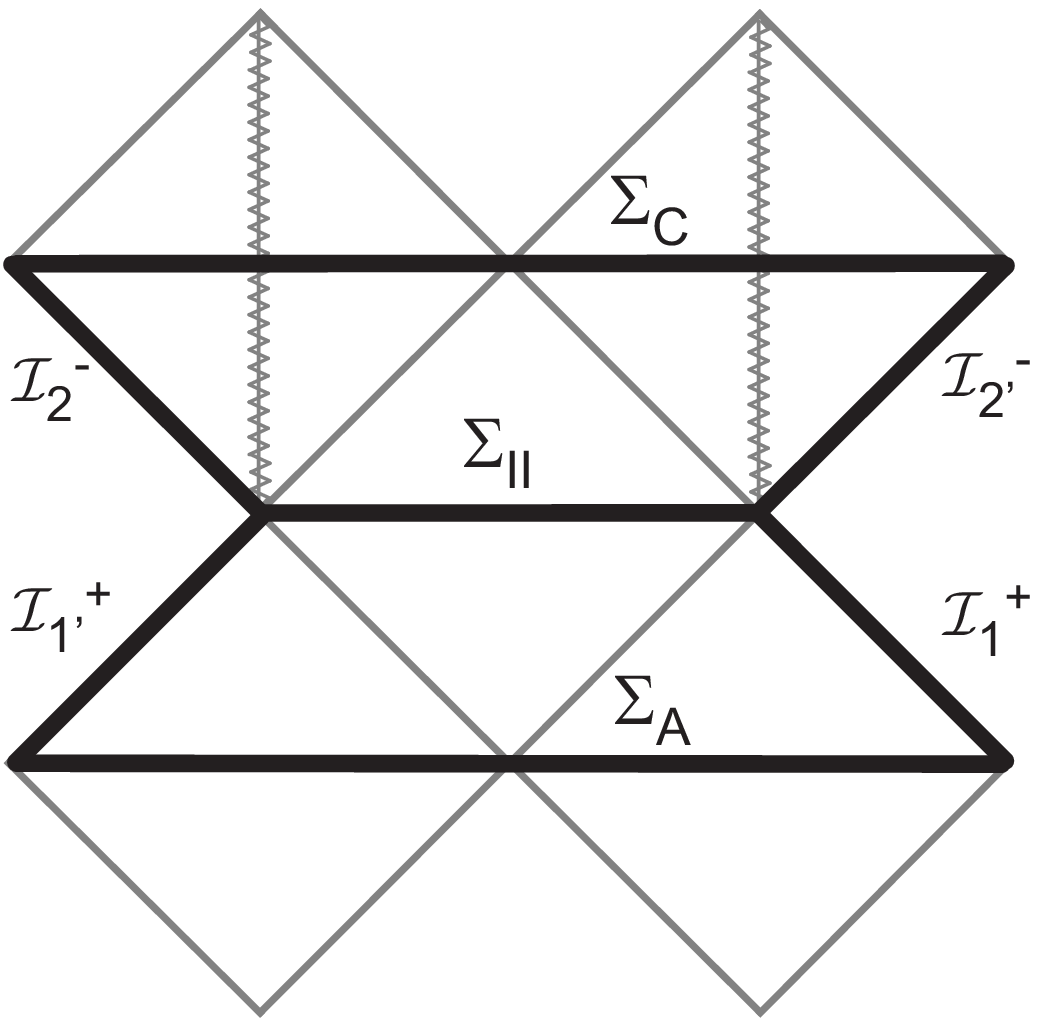}{6 truecm} \figlabel\scray
  Since the regions ${\cal I}_{1}^{+}$, ${\cal
I}_{1'}^{+}$ are causally disconnected and disjoint from the
regions ${\cal I}_{2}^{-}$, ${\cal I}_{2'}^{-}$ the corresponding
restrictions of operators $\phi(x,t)$ and $\Pi(x,t)$ commute.
Therefore, by \nicef\ and \KGcons\ we have
 \eqn\Acomm{ [(\phi, F_{\omega})_{\Sigma_{A}},
(\phi, \bar F_{\omega})_{\Sigma_{C}}] =(F_{\omega},
 F_{\omega})_{\Sigma_{II}} \, .}
The mode $F_{\omega}$ can be chosen to be any arbitrary Killing
vector eigenmode in region II, in particular it can be chosen in
such a way that $(F_{\omega}, F_{\omega})_{\Sigma_{II}}\ne 0$ (an
example of such a Killing eigenmode is the one which vanishes on
the outer horizon of region 1). We conclude that the quantity on
the right hand side of \Acomm\ is a $c$ number which generically does
not vanish. This presents an obstruction to the existence of an
analog to the Hartle-Hawking vacuum \hh\ because there cannot be a
state simultaneously annihilated by the operators $(\phi,
F_{\omega})_{\Sigma_{A}}$ and $(\phi, \bar
F_{\omega})_{\Sigma_{C}}$.~\foot{It 
should be emphasized that the conclusion is drawn in the
framework of semiclassical quantization that does not take
significantly into account any stringy effects.}

Finally, to build such a function as $F_{\omega}$ we start with an
arbitrary Killing vector eigenmode $F_{\omega}^{II}$ in region II.
The values of $F_{\omega}^{II}$ are thus specified on the future
part of the outer horizon and on the past part of the inner
horizon. Consider the restrictions $F_{\omega}^{II}(U_{out})$,
$F_{\omega}^{II}(V_{out})$ to the future half of the outer horizon
$U_{out}>0$, $V_{out}>0$, where $U_{out}$, $V_{out}$ are the
affine parameters. The positive frequency extensions of
$F_{\omega}^{II}$ to the past outer horizon then read \Unruh\
 \eqn\incoma{F_{\omega}(U_{out}) =
\left\{ \matrix{ F_{\omega}^{II}(U_{out}) & \quad U_{out}>0 \cr
e^{\pi
 \omega} F_{\omega}^{II}(-U_{out}) & \quad U_{out}<0 }\right.}
and
 \eqn\incomb{F_{\omega}(V_{out}) = \left\{ \matrix{
F_{\omega}^{II}(V_{out}) & \quad V_{out}>0 \cr e^{\pi
 \omega} F_{\omega}^{II}(-V_{out}) & \quad V_{out}<0 \, .}\right.}
Once the function is specified on all parts of the outer horizon
it has a unique extension to region $A$. Analogously, we extend
$F_{\omega}^{II}$ to region $C$, but this time to consist of only
negative modes with respect to the affine parameters $U_{in}$,
$V_{in}$ on the inner horizon.

%%%%%%%%%%%%%%%%%%%%%%%%%%%%%%%%%%%%%%%%%%%%%%%%%%%%%%%%%%%%%%%%%%%%%%
%%%%%%%             Q U A N T I Z A T I O N    2                 %%%%%
%%%%%%%%%%%%%%%%%%%%%%%%%%%%%%%%%%%%%%%%%%%%%%%%%%%%%%%%%%%%%%%%%%%%%%

\subsec{Quantization 2}
The second approach to quantization of a scalar field in the 2-d CBH
background is
based on the $SL(2, \IR)$ structure underlying the model at hand and
is more inspired by the underlying
CFT and  string theory.
Every mode of the scalar field corresponds to a low lying vertex operator in
string theory. Vertex operators in
${\widetilde{SL}(2, \IR)\times U(1)\over U(1)}$ are vertex operators in
$\widetilde{SL}(2, \IR)\times U(1)=AdS_3\times U(1)$
that are invariant under the gauge transformation
\gaugeact. In string theory we associate different
space-time fields with different modes of the string
(vertex operators) that do not mix under the global
symmetry group (which is the space-time isometry
group). Thus each space-time field is characterized by a unitary
representation of the global symmetry group.
The global symmetry group of the string moving
on a group manifold is the group  itself, in our
case it is $\widetilde{SL}(2, \IR)\times U(1)$.
After doing a KK reduction along the $U(1)$, gauge
invariant vertex operators in $\widetilde{SL}(2, \IR)\times U(1)$
with different momentum along the $U(1)$
factor give different KK particles.
Here however we are interested only in the low lying excitations of
the KK tower. Thus we are interested in the unitary representations of the
group $\widetilde{SL}(2, \IR)$
and the corresponding wave functions which we now review.

Any representation of $\widetilde{SL}(2, \IR)$ is characterized by
two Casimir eigenvalues $j$ and $\epsilon$. The Casimir eigenvalue
$\epsilon$ takes values in the range $[0,1/2]$ and describes how
the states change when we go from one region to another 
in $\widetilde{SL}(2,\IR)$.~\foot{More precisely, 
if we represent points
$g\in\widetilde{SL}(2, \IR)$ as we did in \slparam, \gparam\ then
$\epsilon$ describes how the states change from the point $g$ to
$g\cdot i\sigma_2$ (which is a step along the compact direction of
$SL(2, \IR)$ that is unwrapped in $\widetilde{SL}(2, \IR)$).}
%from one region to another by
%a step of $i\sigma_2$ in $\widetilde{SL}(2, \IR)$.
% how the states change when going from one copy of $PSL(2, \IR)$ to
%another one within $\widetilde{SL}(2, \IR)$.
The second Casimir eigenvalue $j$ is related to the Laplacian
eigenvalue $c_2=-J_3^2+J_1^2+J_2^2$ as $c_2=-j(j+1)$ (where
$J_1,J_2$ and $J_3$ are the generators of $SL(2, \IR)$). For each
representation $\rho$ of $\widetilde{SL}(2, \IR)$ one can define a
wave function on the group manifold as $f(g)=\langle\psi |\rho(g)|
\psi'\rangle$ where $g\in \widetilde{SL}(2, \IR)$. A complete
basis of functions in ${\cal L}^2\(\widetilde{SL}(2, \IR)\)$ is
composed in this way out of  two types of unitary representations.
These two types  are discrete representations, for which $j>-1/2$
take real values, and continuous representations, for which
$j=-1/2+is,\ s\in\IR$. In the discrete representations the wave
function falls exponentially at infinity and describes a
normalized excitation of the field which lives far from the
boundary. In the continuous representations the wave function is
delta function normalizable and describes a scattering mode. While
$j$ has to do with a local behavior of the wave function,
$\epsilon$ is a global characteristic and plays a role only when
we go between regions in $\widetilde{SL}(2, \IR)$.
%a step of $i\sigma_2$ in $\widetilde{SL}(2, \IR)$.
%from one copy of $PSL(2, \IR)$ to the next one in
%$\widetilde{SL}(2, \IR)$.

When we gauge a subgroup inside $\widetilde{SL}(2, \IR)$ only a
gauge invariant subspace of wave functions for  each
representation gives the wave functions  on the coset \foot{For
generic charge the discrete representations do not survive the
Euclidean gauging.}. Note that given two waves with the same
$j$, an observer that sits in a single region cannot tell if the
waves  are of the same type (have the same $\epsilon$) or of
different types (have different $\epsilon$'s). 

After going to the
CBH coset, from the point of view of quantization 1, described in
the previous subsection, picking a specific $\epsilon$ corresponds
to picking a specific one-to-one correlation between the initial
data on $\Sigma_A$ and the initial data on ${\cal I}_{2}^{-}$,
${\cal I}_{2'}^{-}$ (see Fig. \scray).~\foot{In that case the
operators $\phi(x,t)$ and $\Pi(x,t)$ restricted to ${\cal
I}_{1}^{+}$, ${\cal I}_{1'}^{+}$ and to ${\cal I}_{2}^{-}$, ${\cal
I}_{2'}^{-}$ do not commute.} A mode characterized by a fixed
$\epsilon$ that is made from purely positive frequencies with
respect to the affine parameters along the  outer horizon is
necessarily made from a mix of positive and negative frequencies
with respect to the affine parameters along the inner horizon (see
appendix C for details in the case $\epsilon=0$).
%This means that there is no preferred state analog of Hartle-Hawking
%vacuum. In any Fock vacuum the expectation value of the stress
%tensor diverges on, at least, one of the horizons and a large back
%reactions should be considered.
We thus conclude that in the second approach to quantization for a
given species of waves (characterized by a fixed $\epsilon$)
there is no state in which one has thermal distributions in
regions of type A and C with the corresponding temperatures -- a
would be analog of the Hartle-Hawking vacuum for the whole
extended space-time at hand does not exist.

\subsec{Conclusions}

In \KayWald\ Kay and Wald  proved that the Hartle-Hawking vacuum
is the unique vacua which is invariant under the isometries and in
which the expectation value of the stress-energy tensor is regular
on the horizon. This leads us to the conclusion that in any Fock
vacuum of the extended CBH, which is invariant under the
isometries, the expectation value of the stress-energy tensor
diverges at least on one of the horizons. The divergence of the
stress-energy tensor at a smooth point in space-time means that
the energy density there is large and a back reaction has to be
taken into account already in a vacuum state. The singularity is
not a smooth part of the background. If a proper analog of the
Hartle-Hawking vacuum $|\widetilde{HH}\rangle$ with a non-singular
stress-energy tensor on both of the horizons existed, we still
would have to study the behavior of the stress-energy tensor on the
singularities. However, as we showed above, a simple would be
Hartle-Hawking state does not exist and the stress-energy tensor 
is already singular at one of the horizons.

A possible interpretation of this result is the following.
Even in a ground state, particle creation  
beyond the event horizon leads to a large back reaction causing 
a divergent stress-energy tensor at the inner horizon.

%%%%%%%%%%%%%%%%%%%%%%%%%%%%%%%%%%%%%%%%%%%%%%%%%%%%%%%%%%%%%%%%%%%%%%%%%%%
%%%%%%%%%%%%%%                 NAPPI-WITTEN
%%%%%%%%%%%%%%%%%%%%%%%%%%%%%%%%%%%%%%%%%%%%%%%%%%%%%%%%%%%%%%%%%%%%%%%%%%%

\newsec{Cosmology with whiskers}

\subsec{Lorentzian Nappi-Witten solution -- a review}

A coset CFT which is closely related to the 2-d CBH is the cosmological
Nappi-Witten background \NW. Recently, it was studied
in more detail in \refs{\BBBC,\removing}. This background is obtained
from a coset CFT:
 \eqn\nwcoset{\(\widetilde{SL}(2,\IR)\times SU(2)\)/ \(U(1)\times
 U(1)\) \, .}
Let $(g_{1}, g_{2})\in \widetilde{SL}(2, \IR)\times SU(2)$. The
non-anomalous $U(1)\times U(1)$ group action is chosen as
 \eqn\gdefor{\eqalign{
\delta g_{1} &=\epsilon \sigma_{3}g_{1} + (\tilde \epsilon
\cos(\alpha) -\epsilon  \sin(\alpha))g_1 \sigma_3 \, , \cr
%%%%%sign of alpha changed w.r.t. NW paper
\delta g_2 & = i\tilde \epsilon \sigma_{3} g_{2} + (\tilde
 \epsilon \sin(\alpha) + \epsilon \cos(\alpha) )g_{2}i\sigma_3~,
}}
where $\alpha$ is an angular variable analogous to the parameter
$\psi$ that labels the CBH backgrounds. The four-dimensional
space-time manifold described by the corresponding gauged WZW
contains six regions depicted in Fig. 7 (see \BBBC\ for details),
which are cyclically repeated in the maximally extended solution.
Two out of these six regions are time-dependent and describe a
universe evolving from a big bang to a big crunch singularity
(these regions are marked by the letter $C$ in Fig. 7). The remaining
four regions are static, contain closed time-like curves and are
usually referred to as ``whiskers" \BBBC\ (marked by the letter $W$ in
Fig. 7). In addition to the presence of closed time-like curves
the whiskers contain a time-like singularity surface called a
``domain wall'' in \BBBC.
 \fig{Nappi-Witten Cosmology.}{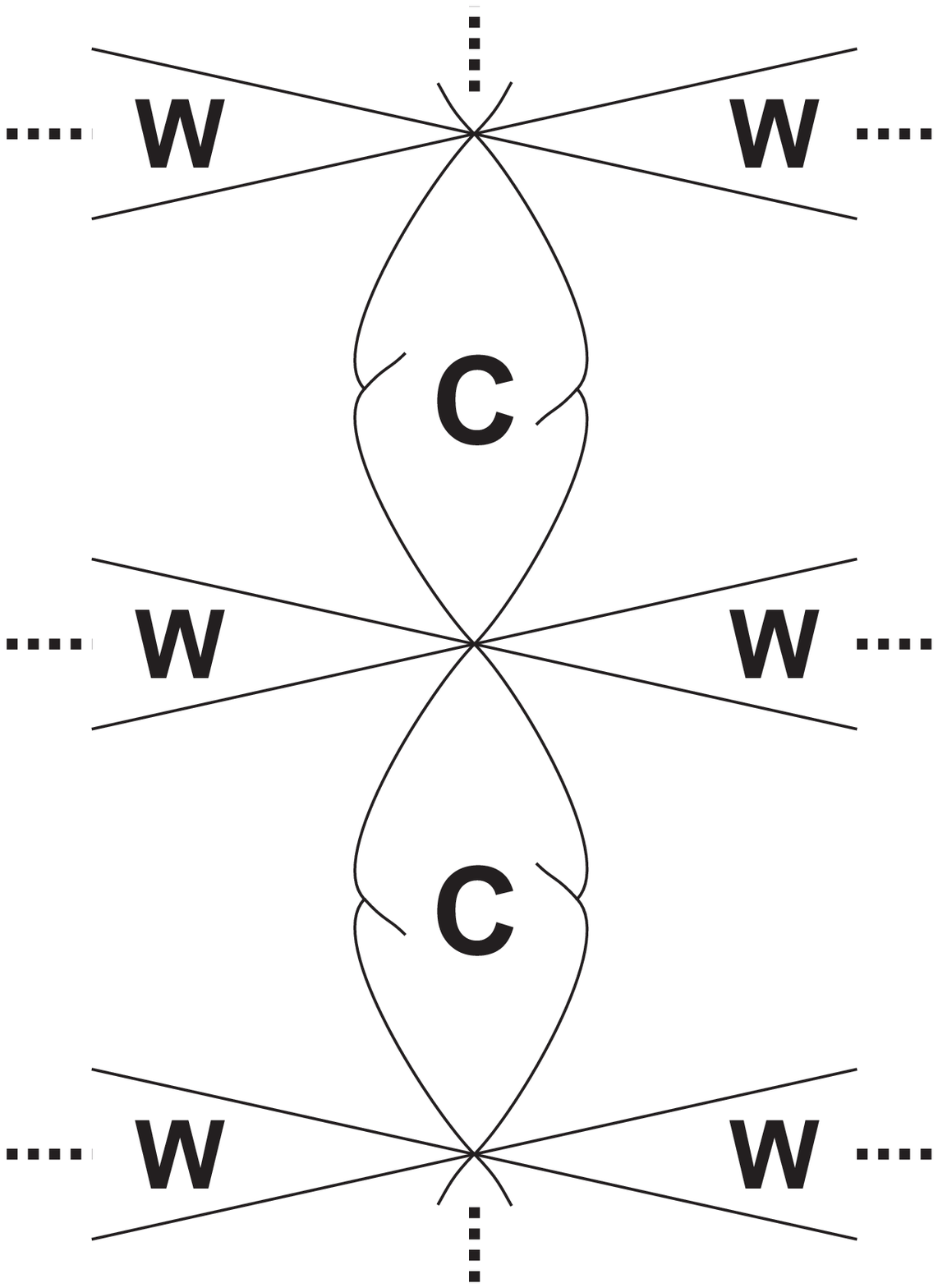}{6 truecm}
 \figlabel\NW
Explicitly the background in a whisker is described as follows.
Let us choose a parameterization of the corresponding submanifold
of $\widetilde{SL}(2, \IR)\times SU(2)$ as
 \eqn\gonetwo{\eqalign{
    g_{1}=&e^{\gamma \sigma_{3}}e^{\theta \sigma_{1}}
    e^{\beta \sigma_{3}},\cr g_{2}=&e^{i\gamma'\sigma_{3}}
    e^{i\theta'\sigma_{2}}e^{i\beta'\sigma_{3}}}}
    (this corresponds to
the $\epsilon=\epsilon'=0$, $\delta=I$ region in the notation of
\BBBC). Here $g_{1}\in \widetilde{SL}(2, \IR)$, $g_{2}\in SU(2)$.
We choose the gauge fixing condition $\gamma=\beta=0$. There are
no residual gauge transformations in this case since $\gamma$ and
$\beta$ are noncompact. The metric  in such a whisker can be
derived to be
 \eqn\Lmetricone{
\frac{ds^{2}}{k} = (d\theta)^{2} + (d\theta')^{2} + g_{\lambda_{+}
\lambda_{+}}(d\lambda_{+})^{2} +
 g_{\lambda_{-}\lambda_{-}}(d\lambda_{-})^{2}\, , }
 \eqn\Lmetrictwo{\eqalign{
g_{\lambda_{-}\lambda_{-}} =&-\frac{ \tanh^{2}(\theta)}{b^{2} -
\tanh^{2}(\theta)\cot^{2}(\theta')} \, , \cr
g_{\lambda_{+}\lambda_{+}}=&\frac{b^{2}\cot^{2}(\theta')}{b^{2}
-\tanh^{2}(\theta) \cot^{2}(\theta')}\, ,  \cr b^{2}=&\frac{
 1-\sin(\alpha)}{1+ \sin(\alpha)} \, .}}
Here $ \lambda_{\pm} =\gamma'\pm \beta' $ have periodicity of
$2\pi$. As evident from the form of the metric, shifts of the
coordinates $\lambda_{\pm}$ generate two commuting isometries.

In addition there are nontrivial B-field and dilaton backgrounds
 \eqn\D{\eqalign{
B_{\lambda_{+}\lambda_{-}}=&\frac{kb^{2}}{b^{2} -
\tanh^{2}(\theta) \cot^{2}(\theta')} \, ,\cr \Phi=&\Phi_{0}
-\frac{1}{2}\log(\cosh^{2}(\theta)\sin^{2}(\theta') -
 b^{2}\sinh^{2}(\theta) \cos^{2}(\theta')) \, .}}
The surface specified by the equation
$$
b^{2} - \tanh^{2}(\theta) \cot^{2}(\theta')=0
$$
is a curvature singularity to which we refer to as  a singular
domain wall.

\subsec{Euclidean Nappi-Witten background} The Euclidean
Nappi-Witten background is obtained as follows. Start with the
$\widetilde{SL}(2, \IR)\times SU(2)$ WZW model and gauge away one
time-like and one space-like $U(1)$ so that the remaining space is
a four-dimensional Euclidean one. A general non-anomalous
$\widetilde{U}(1)\times U(1)$ action of this kind has the form
 \eqn\EgTr{\eqalign{\delta g_{1} =&\epsilon i\sigma_{2}g_{1} + (\tilde
\epsilon \eta_1\eta_2 \sinh(\alpha_{E}) + \epsilon \eta_1
\cosh(\alpha_{E}))g_1 i\sigma_2 \, , \cr \delta g_2 =&i\tilde
\epsilon \sigma_{3} g_{2} + (\tilde \epsilon
\eta_{2}\cosh(\alpha_{E}) + \epsilon \sinh(\alpha_{E})
 )g_{2}i\sigma_3\, .}}
Here $\eta_{1}, \eta_{2}=\pm 1$ are discrete parameters
corresponding to the axial/vector choices of gaugings on
$\widetilde{SL}(2, \IR)$ and $SU(2)$, respectively.

Let us concentrate on the case where $\eta_{1}=1$,
$\eta_{2}=-1$.~\foot{The signs of the $\eta$'s are interchanged by
T-duality.}
We choose parameterizations
 \eqn\gpapametri{\eqalign{
g_{1}=& e^{i\gamma \sigma_{2}} e^{\theta \sigma_{3}}e^{i\beta
\sigma_{2}} \, , \cr g_{2}
=&e^{i\gamma'\sigma_{3}}e^{i\theta'\sigma_{2}}e^{i\beta'\sigma_{3}}
 \, .}}
Here $(\gamma+\beta)$ is the compact time in $SL(2, \IR)$ that
should be unwrapped to obtain the universal cover
$\widetilde{SL}(2, \IR)$. We can perform an initial gauge fixing
by imposing the condition $\gamma=\beta=0$. We obtain the
following metric, B-field and dilaton:
 \eqn\Emetric{\eqalign{
    &\frac{ds^{2}}{k} = (d\theta)^{2} + (d\theta')^{2} + \frac{
    \tanh^{2}(\theta)(d\lambda_{-})^{2}} {b^{2}_{E} +
    \tanh^{2}(\theta)\cot^{2}(\theta')} + \frac{
    b^{2}_{E}\cot^{2}(\theta')(d\lambda_{+})^{2}}{b^{2}_{E} +
    \tanh^{2}(\theta)\cot^{2}(\theta')}  \, , \cr
    &B_{\lambda_{+}\lambda_{-}}=\frac{kb_E^{2}}{b_E^{2}+\tanh^{2}(\theta)
    \cot^{2}(\theta')}~, \cr
    &\Phi= \Phi_{0} -\frac{1}{2}\log(\cosh^{2}(\theta)\sin^{2}(\theta')
    + b_E^{2}\sinh^{2}(\theta) \cos^{2}(\theta'))~,}}
where
 \eqn\bEDef{b^{2}_{E}=\coth^{2}\left(\frac{\alpha_{E}}{2}\right)~,}
and as in \Lmetricone\ $\lambda_{\pm} =\gamma'\pm \beta'$ both have
periodicity of $2\pi$, but now in contrast with the Minkowski
signature case, the coordinate $\gamma -\beta$ is compact. It is
easy to derive from \EgTr\ that the residual gauge transformations
preserving $\gamma+\beta$ and shifting $\gamma-\beta$ by an
integer multiple of $2\pi$ are generated by the shift
 \eqn\resg{\lambda_{-} \to   \lambda_{-} + 2\pi b_{E} \, .}
 Comparing \Emetric\ with \Lmetricone\
we see that the domain wall has disappeared. The subspace
$\theta=\theta'=0$ is singular, it has a trumpet-like curvature
singularity. There are also potential conical singularities at
$\theta=0$, $\theta'=0$ and $\theta'=\pi/2$ subspaces when the
coordinate $\lambda_{-}$ or $\lambda_{+}$ shrinks to zero size.
The identifications $\lambda_{\pm} \sim \lambda_{\pm} + 2\pi$ and
\resg\ are precisely those needed to avoid those conical
singularities. Since the projection with respect to the
transformations \resg\ happens in addition to the original
periodicity of $\lambda_{-}$ of $2\pi$ we see that unless the
parameter $b_{E}$ is rational the coordinate $\lambda_-$ is
completely degenerate.

%%%%%%%%%%% SPECTRUM
Another way to see the same effect is by studying the spectrum. An
irreducible representation of the universal cover of
$SL(2,\IR)\times SU(2)$ is labelled by numbers $m, \bar m, j$,
$m', \bar m', j'$. Even on the universal cover the numbers $m,
\bar m$ must satisfy $ m-\bar m \in {\rm\bf Z}$. The gauge
invariance condition reads
 \eqn\mbarm{\eqalign{
    &m + \bar m \cosh(\alpha_{E}) + \bar m'\sinh(\alpha_{E}) = 0\, ,
    \cr
    &m' - \bar m\sinh(\alpha_{E}) - \bar m' \cosh(\alpha_{E}) =
    0 \, . }}
From these formulas it is easy to derive
$$
    m-\bar m = \frac{1+\cosh(\alpha_{E})}{\sinh(\alpha_{E})} (\bar m'
    - m') \, .
$$
Since $\bar m' - m'$ is an integer we see that $m-\bar m$ can be a
nonzero integer only if
$$
    \frac{1+ \cosh(\alpha_{E})}{\sinh(\alpha_{E})} =
    \coth\left(\frac{\alpha_{E}}{2}\right) = b_{E}
$$
is a rational number. Generically however we have only the zero
solution $m=\bar m=0$ that describes only zero momentum states.
%%%%%%%%%%%%%%%%%%%%%%%%%%%%%

The nature of this phenomenon can be tracked down to having a
situation when a compact torus is quotiented by a $U(1)$ whose
position is incommensurate with the torus lattice. The
corresponding gauge orbit is everywhere dense on the torus and we
are left with no momenta (except zero) on the quotient.

In the case when $b_{E}$ is rational the corresponding coset space
is a four-dimensional Euclidean manifold that can be described as
follows. Assume that $b_{E}=P/Q$ where $P$, $Q$ are relatively
prime. Then it follows from \resg\ that $\lambda_{-}$ has a radius
of $\frac{1}{Q}$ and $\lambda_-/b_E$ has a radius of
$\frac{1}{P}$. This implies then that the metric \Emetric\ has an
orbifold type conical singularities respectively at $\theta'=0$ of
a deficit angle $2\pi/Q$ and at $\theta=0$ of a deficit angle
$2\pi/P$. Thus the only nonsingular case is $|b_{E}|=1$ (in that
case gauging $SL(2, \IR)$ instead of $\widetilde{SL}(2,\IR)$ would
not add a new identification and we would get the same Euclidean
geometry).

In parallel with our discussion of Euclidean CBH background the
Euclidean space \Emetric\ is obtained from the Lorentzian one
\Lmetricone\ by the following Wick rotation:
 \eqn\Wick{\alpha \to   i\alpha_{E}-\frac{\pi}{2}~,}
where we included a shift by $-\pi/2$ to match the conventions of
\Lmetrictwo\ with \bEDef\ (which reads $b^2\to -b_E^2$). From the
point of view of the Wick rotated background the periodicities of
$\lambda_\pm$ are imposed to avoid  conical singularities at
$\theta=0$, $\theta'=0$ and $\theta'=\pi/2$. As in the above
discussion of the Euclidean coset, unless $|b_E|=1$, the required
identifications yield a degenerate, effectively three-dimensional
spectrum.

In the asymptotic region $\theta \to \infty$ the metric \Emetric\
takes the form
$$
    \frac{ds^{2}}{k} \approx (d\theta)^{2} + (d\theta')^{2} + \frac{
    (d\lambda_{-})^{2}} {b^{2}_{E} +\cot^{2}(\theta')}
    +\frac{b^{2}_{E}\cot^{2}(\theta')(d\lambda_{+})^{2}}{b^{2}_{E} +
    \cot^{2}(\theta')}~,
$$
which is uniformly bounded in both $\lambda_{\pm}$ directions. The
maximal asymptotic radius of the $\lambda_{-}$ direction is $b_{E}^{-2}$ and
is achieved at $\theta'=\pi/2$ while the maximal asymptotic radius of the
$\lambda_{+}$ direction is $b_{E}^{2}$ which is achieved at
$\theta'=0$.  The maximal radii are equal when   the conical
singularities are absent that is when $b_{E}^{2}=1$. In the
Minkowski signature space one of the Killing vectors is time-like.
Since in the Euclidean solution the asymptotic values of both
$\lambda_{+}$ and $\lambda_{-}$ (Euclidean) directions are bounded
this suggests that at $b_{E}^{2}=1$ the Euclidean solution defines
a vacuum state in the original Minkowski signature space that is
thermal. The dependence of the asymptotic sizes on $\theta'$
presumably can be interpreted in terms of the high anisotropy of
the outgoing thermal radiation. In the Euclidean space both
Killing directions are completely on equal footing. Thus it seems
likely to us that the Euclidean solution defines a vacuum
characterized by a canonical distribution in the eigenvalues of
both Killing vectors $\frac{\partial}{\partial\lambda_{\pm}}$.

%If $b_{E}$ is integer then unless $|b_{E}|=1$ we have an orbifold
%singularity at $\theta'=0$ and if $1/b_{E}$ is integer then unless
%$|b_{E}|=1$ we have an orbifold singularity at $\theta=0$. The
%only smooth Euclidean geometry is when $|b_{E}|=1$, (in that case
%gauging $SL(2, \IR)$ instead of $\widetilde{SL}(2,
%\IR)$ would not add a new identification and we would have
%got the same Euclidean geometry).

In the above discussion of the Euclidean background we
concentrated on a coset \EgTr\ with $\eta_{1}=1$, $\eta_{2}=-1$, which
is related by the Wick rotation \Wick\ to the Minkowski
signature whisker characterized in the notation of \BBBC\ by the
values $\epsilon=\epsilon'=0$, $\delta=I$. 
Each of the pairs of whiskers (1, 3) and (1', 3') is described by the same
Euclidean background.  The whiskers 1 and 1' are related by interchanging
$\lambda_ +$ with $\lambda_-$ (the same goes for 3 and 3'). The Wick rotated
background in whiskers 2 and 4 (obtained by the same Wick rotation as
in \Wick) can also be obtained by considering the Euclidean coset
\EgTr, corresponding to the choice $\eta_1=-1$, $\eta_2=1$.
The two Euclidean cosets are T-dual to each other. The corresponding metric,
B-field and dilaton are obtained from the ones in \Emetric\ by the change
$\theta \to i\pi/2 - \theta$. The background in 2' (4') is obtained by the
interchange mentioned above. The residual gauge transformations for the T-dual
Euclidean coset are now generated by the shift
$$      
\lambda_{+} \to \lambda_{+} + 2\pi b_{E}^{-1} \, .     
$$
The two types of the Euclidean whiskers obtained have the same singularity
structure outlined before. Thus, for irrational $b_{E}$ they both degenerate.
The conical singularities are absent only when $|b_{E}|=1$ and the trumpet-like
curvature singularity is present for both types.  
The backgrounds in the regions 1 and 2 (as well as 3, 4 and in the  
corresponding primed whiskers) are different for a general value of 
the variable $\theta$. However, for large values of that variable and 
$|b_{E}|=1$ they have the same form.
This is unlike the generic case of the CBH, where the two Euclidean backgrounds
have also two different asymptotic temperatures. 
It is more like the case of the extremal CBH.
%The same Euclidean
%space takes care of the whiskers $1$, $1'$, $3$, $3'$ (see \BBBC)
%with the coordinates $\lambda_{+}$ and $\lambda_{-}$ interchanged
%for the whiskers $1'$ and $3'$. The Wick rotated metric (obtained
%by the same Wick rotation
% \Wick)  of the remaining whiskers is obtained by considering the Euclidean
%coset \EgTr\ corresponding to
%the choice $\eta_1=-1$, $\eta_2=1$.  The two  Euclidean cosets are
%T-dual to each other. The corresponding metric and B-field are
%obtained from the ones in \Emetric\ by a change $\theta \to i\pi/2
%- \theta$. The residual gauge transformations are now generated by
%the shift
%$$
%\lambda_+ \to \lambda_{+} + 2\pi b_{E}^{-1} \, .
%$$
%The T-dual Euclidean coset has the same degeneracy features as the
%original one. For irrational $b_{E}$ it degenerates and the
%conical singularities are absent only when $|b_{E}|=1$.

To summarize, we see that  in distinction with the CBH
case  the Euclidean coset CFT's that are related to the Minkowski
signature CFT via Wick rotation demonstrate a certain degeneracy
or non-smoothness  for generic values of the Euclidean mixing angle
parameter $b_{E}$. For irrational values of $b_{E}$ one coordinate
is degenerate and, classically, the Euclidean background degenerates to a
three-dimensional one. For a rational $b_{E}\ne \pm 1$ the
resulting background is four-dimensional but possesses
orbifold-type conical singularities, leading to a degenerate spectrum.
This situation is very much
reminiscent of the situation with Euclidean Milne Universe \BP.
The corresponding space also generically has a conical singularity
and its spectrum is truncated to zero modes for generic values of a
certain angular parameter that is analogous to our $b_{E}$. In
addition to those features  the Euclidean backgrounds for all
values of $b_{E}$ also contain a trumpet-like singularity. The
last one can be potentially removed by the method developed in
\removing. We leave this question to a future investigation.

\vskip 1cm \noindent {\bf Acknowledgements:} We thank O.~Aharony, J.~Barbon,
S.~Elitzur, N.~Itzhaki, B.~Kol, A.~Ori, A.~Pakman and K.~Skenderis
for discussions. This work is supported in part by the Israel
Academy of Sciences and Humanities -- Centers of Excellence
Program, by GIF -- German-Israel Bi-National Science Foundation,
and the European RTN network HPRN-CT-2000-00122. The work of E.R.
is supported in part by BSF -- American-Israel Bi-National Science
Foundation. The work of A.S. is supported in part by the Horowitz
Foundation.

%%%%%%%%%%%%%%%%%%%%%%%%%%%%%%%%     APPENDIX
%%%%%%%%%%%%%%%%%%%%%%%%%%%%%%%%%%%%%%%%

\appendix{A}{Euclidean Kerr black hole}
For completeness, we show how one can compute the chemical
potential $\Omega_H$ conjugate to angular momentum in a Kerr black hole, 
in an analogous way to the Euclidean computation of
$\mu_{\rm el}$ for RN.

To make the analog more clear we first start by noting that the
2-d CBH is actually a Kaluza-Klein (KK) reduction of a 3-d
dilatonic rotating black string, where the 2-d gauge field is
related to the black string angular momentum. In what follows we
will show how the condition that $A_0$ vanishes at the tip of the
Euclidean CBH (ECBH) comes from the regularity condition of the
Euclidean black string. We then show that the condition for the
regularity of the Euclidean black string is the same as the one
for the Euclidean Kerr black hole, and read $\Omega_H$ from
the regular Kerr metric.

The 3-d black string metric in the region of the black string
which after the KK reduction becomes region A of the CBH is given
by \ChargedBh:
 \eqn\rotbs{
    ds^2=d\theta^2-{dy^2-2pdydx\over\coth^2(\theta)-p^2}+dx^2~,}
where $x=x+2\pi L$. After the KK reduction the 2-d metric, dilaton
and gauge field become as in \twodmetric,\DGA. After Euclidean
continuation $y\rightarrow i\tau$ and $p\rightarrow -ip$, \rotbs\
becomes~\foot{Alternatively, this background is obtained from the
Euclidean coset \gaugeact.}
 \eqn\EclidBs{
    ds^2=d\theta^2+{d\tau^2+2pd\tau dx\over\coth^2(\theta)+p^2}+dx^2~.}
Near the horizon $\theta=0$ of the black string the metric \rotbs\
is
 \eqn\asas{ds^2\sim d\theta^2+\(dx+p\theta^2d\tau\)^2+\theta^2d\tau^2~.}
Upon coordinate transformation $\tilde x=x-c\tau$ (which is
equivalent to a large gauge transformation in the Lorentzian CBH),
the same metric is
 \eqn\LorensBs{
    ds^2\sim d\theta^2+\(d\tilde
    x+(c+p\theta^2)d\tau\)^2+\theta^2d\tau^2~.}
Now, to have a non-singular metric we compactify the $\tau$
direction $(x,\tau)=(x,\tau+2\pi)$. If instead we identify
$(\tilde{x},\tau)=(\tilde{x},\tau+2\pi)$ then at $\theta=0$ we are
identifying $x=x+2\pi L$ and $x=x+2\pi c$, where the metric is
$ds^2=d\theta^2+dx^2$. If $L/c$ is not rational then the geometry
is singular at $\theta =0$. The same singularity is seen in the
ECBH. A KK reduction along the $\tilde{x}$ direction leads to a
singular gauge field at $\theta=0$, unless all the charges are
quantized in units of $1/c$. Upon the KK reduction along the $x$
direction, we get a regular gauge field which supports a
non-trivial Wilson loop at infinity. The chemical potential is
read from this asymptotic Wilson loop. In a Kerr black hole the
situation is the same.

The Kerr metric in four dimensions is given by:
 \eqn\kerrmetric{
    ds^2={\rho^2\over r^2f}dr^2-{r^2f\over\rho^2}\[dt-a\sin^2\theta
    d\phi\]^2+\rho^2d\theta^2 +{\sin^2\over \rho^2}\[(r^2+a^2)
    d\phi-adt\]^2~,}
where
 \eqn\fkerr{
    f=1-{2M\over r}+{a^2\over r^2}\ ,\qquad
    \rho^2=r^2+a^2\cos^2\theta~.}
The horizons are located at ($M>|a|$):
 \eqn\rkerr{r_\pm=M\pm\sqrt{M^2-a^2}~.}
The Euclidean metric is obtained by taking $t\rightarrow i\tau$
and at the same time $a\rightarrow ia_{_E}$:
 \eqn\euclidkerr{
    ds^2_{_E}={\rho^2\over r^2f}dr^2+{r^2f\over\rho^2}\[d\tau-
    a_{_E}\sin^2\theta d\phi\]^2+\rho^2d\theta^2 +{\sin^2\theta
    \over \rho^2}\[(r^2-a_{_E}^2) d\phi+a_{_E}d\tau\]^2~.}
In terms of $\tilde{r}\equiv r-r_+$ the Euclidean metric near the
horizon becomes:
 \eqn\euclidkerratH{\eqalign{
    ds^2_{_E}\sim &{r_+^2-a_{_E}^2\over \tilde{r}(r_+-r_-)}d\tilde{r}^2+
    {\tilde{r}(r_+-r_-)\over r_+^2-a_{_E}^2}\[d\tau-a_{_E}\sin^2\theta
    d\phi\]^2 \cr +& (r_+^2-a_{_E}^2)d\theta^2 +\sin^2\theta (r_+^2-
    a_{_E}^2)\[d\phi+\Omega_Hd\tau\]^2~,}}
where
 \eqn\omegakerr{\Omega_H={a_{_E}\over r_+^2-a_{_E}^2}}
is the angular velocity at the horizon. This Euclidean geometry is
singular at $\theta =0,\ \tilde{r}=0$ since at $\theta=0$
\euclidkerratH\ becomes
 \eqn\Ekerr{
    ds^2_{_E \,{\displaystyle\vert}_{\theta=0}}\sim {r_+^2-a_{_E}^2\over
    \tilde{r} (r_+-r_-)}d\tilde{r}^2+ {\tilde{r}(r_+-r_-)\over
    r_+^2-a_{_E}^2}d\tau^2~.}
To smooth out the geometry we compactify the Euclidean time
 \eqn\taukerr{\tau=\tau+{2\pi\over\kappa}=\tau+2\pi\beta~,}
where the surface gravity $\kappa$ is given by
 \eqn\kappakerr{\kappa={1\over 2}\({\tilde{r}(r_+-r_-)\over
    r_+^2-a_{_E}^2}\)'_{\,{\displaystyle\vert}_{\tilde{r}=0}}=
    {r_+-r_-\over 2(r_+^2-a_{_E}^2)}~.}
Comparing \EclidBs\ with \euclidkerratH\ at $\tilde{r}=0$ and
$\theta\sim 0$, we see that in order to avoid a singularity at
$\theta=0$ we should add to \taukerr\ a shift in
$\tilde{\phi}=\phi+\Omega_H\tau$ as follows:
 \eqn\phikerr{(\tilde{\phi},\tau)=(\tilde{\phi},\tau+2\pi\beta)~.}
Now asymptotically the metric \euclidkerr\ becomes
 \eqn\euckerr{
    ds^2_{_E}\sim dr^2+d\tau^2+r^2d\theta^2+
    r^2\sin^2\theta\(d\tilde{\phi} -\Omega_Hd\tau\)^2~.}
As in the 3-d black string case, upon the KK reduction along the
$\tilde{\phi}$ coordinate we get a non-trivial Wilson loop at
infinity from which we read off the chemical potential for charged
particles $\mu_{\rm el}=\Omega_H$. Before the KK reduction this charged
particles are particles with momentum along the $\tilde{\phi}$
direction which are particles with angular momentum. We learn that
in a Kerr black hole the chemical potential conjugate to angular
momentum is $\Omega_H$, which is by now very well known.

\appendix{B}{Relations between modes in different regions}
In this appendix we discuss how to extend wave functions from one
region to a basis in the extended CBH background, while preserving
analyticity with respect to the affine parameters on the horizons.
This basis of wave functions can be used to bring correlators in
any state between operators in different regions to a linear
combination of correlators where all insertions are in the same
region with some imaginary shifts. We start by reviewing a simpler
example of an analogous construction in Rindler space. Rindler
space is a flat Minkowski space-time as seen by an observer that
travels with a constant acceleration. Although, as seen by the
accelerated observer, there are parts of the space-time which are
hidden behind horizons, there is noting special happening at those
points physics-wise. A vacuum state (for example, the Minkowski
vacuum) in which there is noting special at those points (the
expectation value of the stress-energy tensor is everywhere
smooth) is built out of modes which are locally analytic with
respect to any locally flat coordinate system (for example, the
affine parameter along the bifurcate horizons near the bifurcating
pint).
 \fig{Rindler coordinate patches.}{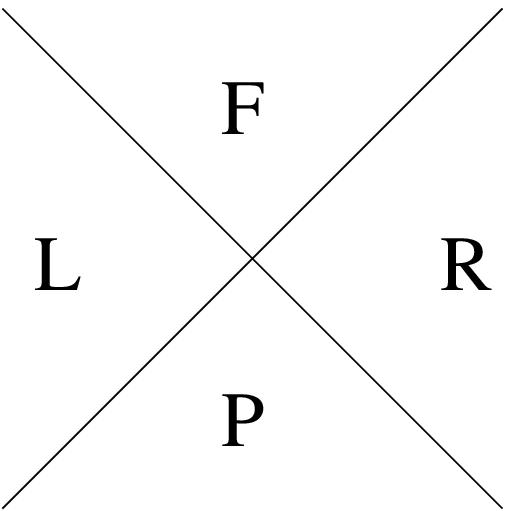}{3 truecm}
 \figlabel\Rin
Rindler coordinates ($r,\eta$) run from $-\infty$ to $\infty$
and cover Minkowski space-time ($x,t$) in
four patches (see Fig. \Rin),
 \eqn\Rindler{x\pm t=\left\{\matrix{re^{\pm\eta} & R,L\cr
\pm re^{\pm\eta} &F,P}
 \right.\ ,\qquad ds^2=dx^2-dt^2=\left\{\eqalign{-r^2d\eta^2+dr^2 &
\quad R,L \cr
 r^2d\eta^2+dr^2&\quad F,P}\right.}
where in regions R and F we have $r>0$ and $r<0$ in regions P and
L. The region R is obtained from the region F by a transformation
$(r,\eta)\rightarrow (\mp ir,\eta\pm i{\pi\over 2})$, where the
line separating the two regions is invariant. In the same way the
region L is obtained from the region F by $(r,\eta)\rightarrow
(\pm ir,\eta\pm i{\pi\over 2})$. The modes which are locally
analytic on the horizons have their values in the regions L and R
obtained from their value in region F by the above
transformations.

The situation on each bifurcate horizon of the CBH background is
the same. Here we parameterize the regions in $SL(2, \IR)$
correspond to the compact regions of the CBH by $g_{II}=e^{{1\over
2}(z+y)\sigma_{3}}e^{i\theta\sigma_2}e^{{1\over
2}(z-y)\sigma_{3}}$ where $\theta\in \[-\pi/2,0\],\
\[0,\pi/2\],\ \[\pi/2,\pi\]$ in regions I, II and III
correspondingly. We fix the gauge by setting $z$ to zero. In that
parametrization the background fields of regionz I, II and III are
obtained from \twodmetric\ by taking $\theta\rightarrow i\theta$.
The CBH metric in the $(y,\theta)$ coordinates (regions 1, 1', I,
II) or $(y/p^2,\theta)$ coordinates (regions 2, 2', II, III) looks
locally near the bifurcating points as a Rindler metric and the
dilaton is approximately constant. The analog of the coordinates
$x\pm t$ in the Rindler space are the affine parameters $U,V$
along the horizons. The local coordinates $(y,\theta)$ are given
in terms of the affine parameters by
%{\bf (in section 1 we have to add the
%parametrization and metric of the compact regions (SSS).)
%WHY NOT HERE ???}
\eqn\affineparam{V_A\pm U_A=\left\{\matrix{(e^y\mp e^{-y})\sh(\theta) &
1,1'\cr
(e^y\pm e^{-y})\sin(\theta) &I,II} \right.\ ,\qquad V_C\pm
U_C=\left\{\matrix{(e^{y/p^2}\mp e^{-y/p^2})\sh(\theta) & 2,2'\cr
 (e^{y/p^2}\pm e^{-y/p^2})\cos(\theta) &II,III} \right.~.}
Regions 1, 1', 2 and 2' are obtained from region II by~\foot{This
coordinate transformations descend from the parameterizations
of the corresponding regions in $SL(2, \IR)$.}
 \eqn\analyticcon{\eqalign{(y,\theta)_{II}\qquad\longrightarrow &
\left\{\eqalign{
\quad (y\pm i\pi/2,\mp i\theta)\quad\ &1 \cr\quad (y\pm i\pi/2,\pm
i\theta)\quad\ &1' }\right. \cr
(y/p^2,\pi/2-\theta)_{II}\longrightarrow &\left\{\eqalign{
(y/p^2\pm i\pi/2,\mp i\theta)\quad &2 \cr (y/p^2\pm
 i\pi/2,\pm i\theta)\quad &2' }\right.}}
Now one can start from any mode in region II and use relations
\analyticcon\ to define the modes in regions 1, 1', 2 and 2'. The
global modes one obtains are analytic on the horizons\foot{To be
more precise, the modes have branch cuts in the complex planes so
one has to specify exactly how to go from one point to the other.
For example the modes in region 1 are obtained from a mode in
region II by the analytic continuation $(\theta,y)(\gamma)=
(\theta e^{-i{\pi\over 2}\gamma},y+i\gamma\pi/2)$, where $\gamma$
is a real parameter that runs from 0 to 1.}. If for example one
defines a mode in region 1 by the transformation
$(y,\theta)_{II}\rightarrow (y+ i\pi/2,-i\theta)$
($(y,\theta)_{II}\rightarrow (y- i\pi/2,i\theta)$) then this mode
will be made from purely positive (negative) frequencies with
respect to the affine parameter $U_A$.

If we properly use this modes to quantize the field separately in
regions 1 and 1' or in regions 2 and 2' we will observe the HH
vacuum of those regions. But, as proved in subsection (2.3), such
a global Fock vacuum does not exist.

\appendix{C}{Hartle-Hawking quantization with
$\epsilon=0$}

In this appendix we give some details regarding  the quantization
of a scalar field from the point of view of section 3.2
(quantization 2). A scalar field at hand is characterized by a
representation of the parent $\widetilde{SL}(2, \IR)$ which is
labelled by the Casimir eigenvalues  $j$ and $\epsilon$. Here we
consider only the continuous representations with $\epsilon=0$. We
construct the Hartle-Hawking (HH) vacuum in the regions of type A
and show that it does not match with the analog of the HH vacuum
in the regions of type C.
%Since the HH vacuum is the unique state in which the expectation
%value of the stress tensor is regular in the corresponding
%regions,\foot{This is true for any
%$\epsilon$.}
We will conclude that for $\epsilon=0$ there is no Fock
vacuum in which there is a thermal distribution in
regions of type A and C with their corresponding temperatures.
The generalization for any $\epsilon$ is straightforward.

A basis of wave functions in region 1 which vanish in region 1'
and which are eigenfunctions of the Killing vector $\xi$ in
\killing\ with eigenvalue $\omega$ is given by
 \eqn\modeone{
    K_{-+}(\lambda,\mu;j;g)\ ,\qquad
    K_{-+}(-\lambda-2j,-\mu-2j;j;g)^*~,}
where
 \eqn\omegadef{
    \lambda-\mu=-i{\omega\over\kappa_1}\ ,\qquad
    \lambda+\mu=-2j=1-2is~.}
$K_{\pm\pm}$ are matrix elements of the representation
characterized by $\epsilon=0$ and $j$ (the explicit form of these
matrix elements can be found in \Vil), $\kappa_1$ is the surface
gravity of the outer horizon and we take $\omega,s>0$. These modes
are not orthogonal to each other in the Klein-Gordon norm \KGn. An
orthogonal basis is given by the following linear combinations of
the modes \modeone:
 \eqn\fpzero{\eqalign{
    F^+_{\omega,1}&\propto K_{-+}(\lambda,\mu;j;g)
    + {B(\lambda,1-\mu)\over B(-\lambda-2j,1+\mu+2j)^*}
    K_{-+}(-\lambda-2j,-\mu-2j;j;g)^*~, \cr
    F^+_{\omega,2}&\propto K_{-+}(\lambda,\mu;j;g)+
    {B(\lambda,2j+1)+B(-\lambda-2j,2j+1)\over
    B(\lambda,2j+1)^*+B(-\lambda-2j,2j+1)^*}
    K_{-+}(-\lambda-2\tau,-\mu-2\tau;j;g)^*~.}}
These modes are orthogonal since $F^+_{\omega,1}$ vanish at the
past null infinity and $F^+_{\omega,2}$ vanish
at the past horizon and the Klein-Gordon norm can be calculated at the
past infinity of region 1. The
corresponding images of these modes in region 1'
($F^-_{\omega,1}$ and $F^-_{\omega,2}$) are given by
interchanging $K_{-+}$ and $K_{+-}$. 
We define the following linear combinations
 \eqn\Hmodes{\eqalign{
    &H^+_{\omega,i}=\ch(\phi_1) F^+_{\omega,i}+\sh(\phi_1)
    F^-_{\omega,i}~,\cr &H^-_{\omega,i}=\sh(\phi_1)F^+_{\omega,i}
    + \ch(\phi_1) F^-_{\omega,i}~,}}
where $\th(\phi)=e^{-\pi{\omega\over\kappa_1}}$ and $i=1,2$.
According to \refs{\Unruh,\Israel} $H^+_{\omega,i}$
($H^-_{\omega,i}$) have purely positive (negative) frequencies
with respect to the affine parameters along
the bifurcate Killing horizon of regions 1 and 1'.
After the scalar field is quantized the HH vacuum of
regions 1 and 1' is obtained if we associate an annihilation
operator $a^{(+)}_{\omega,i}$ to
$H^+_{\omega,i}$ and a creation operator $\(a^{(-)}_{\omega,i}\)^\dagger$
to $H^-_{\omega,i}$ where $i=1,2$. In
terms of the modes $H^\pm$ the quantum field $\phi$
is given by~\foot{There is no sum over $\omega$ since
different $\omega$'s correspond to different fields.}:
 \eqn\phizero{
    \Phi_\omega(x)=\sum_{i=1}^2\(H^+_{\omega,i}a^{(+)}_{\omega,i}
    +H^-_{\omega,i}\(a^{(-)}_{\omega,i}\)^\dagger +{\rm h.c.}\)~.}
To reproduce an analogous construction in regions 2 and 2'
we start with the modes
 \eqn\modeone{
    K_{++}(\lambda,\mu;j;g)\ ,\qquad
    K_{++}(-\lambda-2j,-\mu-2j;j;g)^*~,}
which are eigenfunctions of the Killing vector with eigenvalues $\omega$.
These modes form a basis of wave
functions in region 2' and vanish in region 2.
In parallel with \Hmodes\ we obtain  purely positive
(negative) frequency  modes $\tilde H^+_{\omega,j}$
($\tilde H^-_{\omega,j}$) relative to the affine
parameters along the inner horizon.  These modes  are given by
linear combinations of \modeone\ and their
images
 \eqn\modeoneimag{
    K_{--}(\lambda,\mu;j;g)\ ,\qquad
    K_{--}(-\lambda-2j,-\mu-2j;j;g)^*,}
with the thermal factor $\th(\phi_2)= e^{-\pi{\omega\over\kappa_2}}$.
Here $\kappa_2$ is the surface
gravity of the inner horizon. For a given $\epsilon$ there is an
($\epsilon$-dependent) non-trivial
Bogolubov transformation between the modes of regions 1 and 1',
and the modes of regions 2 and 2'. To show
that this Bogolubov transformation is not trivial let us examine a mode
 \eqn\yzero{Y^+=\ch(\phi_1) K_{-+}+\sh(\phi_1) K_{+-}~,}
which has purely positive frequencies with respect to the affine
parameters along the outer horizon. Let $U>0$ be the affine
parameter along the Killing horizon between regions 2' and III,
and $U<0$ the affine parameter along the Killing horizon between
regions 2 and II. The restrictions  of $K_{-+}$ and $K_{+-}$ to
the horizon  are
 \eqn\konhoizon{\eqalign{
    K_{-+}(\lambda,\mu;j;U)_{|_U<0}=&{(-1)^{2\epsilon}\over 2\pi
    i}B(1+\mu+2j,-\lambda-\mu-2j)|U|^{-i{\omega\over\kappa_2}}\cr
    K_{-+}(\lambda,\mu;j;U)_{|_U>0}=&{(-1)^{2\epsilon}\over 2\pi
    i}B(\lambda,-\lambda-\mu-2j)U^{-i{\omega\over\kappa_2}}\cr
    K_{+-}(\lambda,\mu;j;U)_{|_U<0}=&{1\over 2\pi
    i}B(\lambda,-\lambda-\mu-2j)|U|^{-i{\omega\over\kappa_2}}\cr
    K_{+-}(\lambda,\mu;j;U)_{|_U>0}=&{1\over 2\pi
    i}B(1+\mu+2j,-\lambda-\mu-2j)U^{-i{\omega\over\kappa_2}}}}
So~\foot{Here $\theta(x)$ is 1 for $x>0$ and 0 otherwise.}
 \eqn\tdef{\eqalign{
    &Y^+(\lambda,\mu;j;U)\propto \cr
    &\theta(-U)|U|^{-i{\omega\over\kappa_2}} +
    {B(\lambda,-\lambda-\mu-2j)+e^{-\pi{\omega\over\kappa_1}}
    B(1+\mu+2j,-\lambda-\mu-2j)\over B(1+\mu+2j,-\lambda-\mu-2j)
    +e^{-\pi{\omega\over\kappa_1}}B(\lambda,-\lambda-\mu-2j)}
    \theta(U)U^{-i{\omega\over\kappa_2}}}}
is not analytic on the lower half of the complex $U$ plane and, therefore,
it contains also negative
frequencies with respect to $U$. We conclude that the Bogolubov
transformation at hand is not trivial and
the corresponding vacuum is not thermal in regions of type C.

\bigskip
\listrefs

\end